\documentclass[10pt,a4paper]{article}
\usepackage[dvips]{color}
\usepackage{epsfig}
\usepackage{amsmath}
\usepackage{graphicx}
\usepackage{authblk}
\usepackage{amssymb,amsmath}
\usepackage{amsmath}
\usepackage{tabularx}

\begin{document}
\title{\bf Interacting extended Chaplygin gas cosmology in Lyra Manifold}
\author{{M. Khurshudyan$^{a}$\thanks{Email:
khurshudyan@yandex.ru, martiros.khurshudyan@mpikg.mpg.de}}\\
$^{a}${\small {\em Max Planck Institute of Colloids and Interfaces,\\ Potsdam-Golm Science Park Am Mhlenberg 1 OT Golm, 14476
Potsdam}}\\ }  \maketitle
\begin{abstract}
The subject of our interest is an extended Chaplygin gas cosmology. In Literature a variety of cosmological models exist studying the behavior of the universe in the presence of the Chaplygin gas. From its initial form Chaplygin gas evolved and accepted different EoS-s and we will work with one of them. The main purpose of this work is to study behavior of the universe in Lyra Manifold with a varying  Effective $\Lambda$-Term, which gives us modified field equations. We are also interested in the behavior of the universe in the case of an existing coupling between the quintessence DE and extended Chaplygin gas. We applied observational constraints and causality issue on our model to separate physically relevant behavior of the phenomenological model.  
\end{abstract}

\section{\large{Introduction}}
\begin{figure}[h!]
 \begin{center}$
 \begin{array}{cccc}
 \includegraphics[width=80 mm]{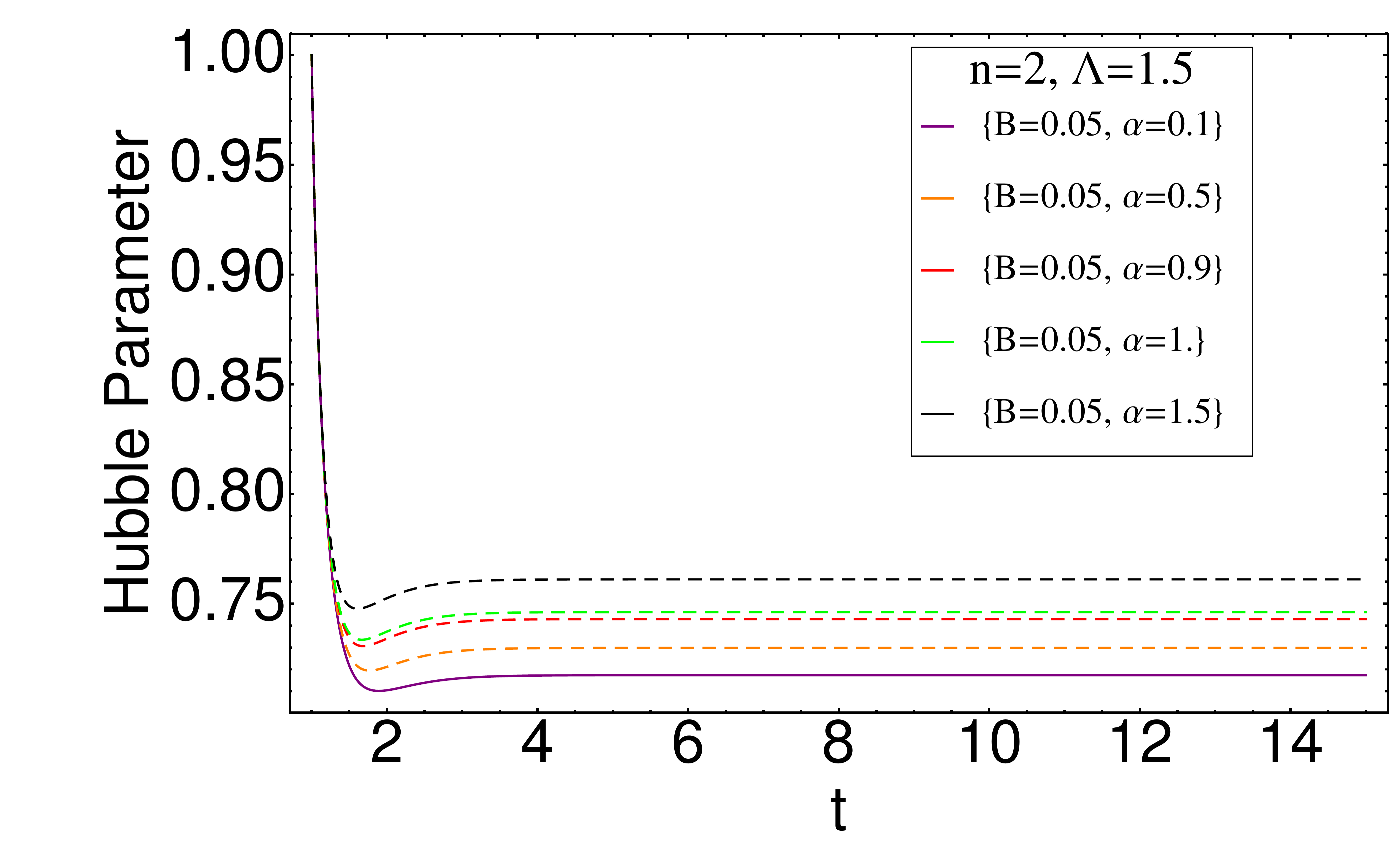} &
\includegraphics[width=80 mm]{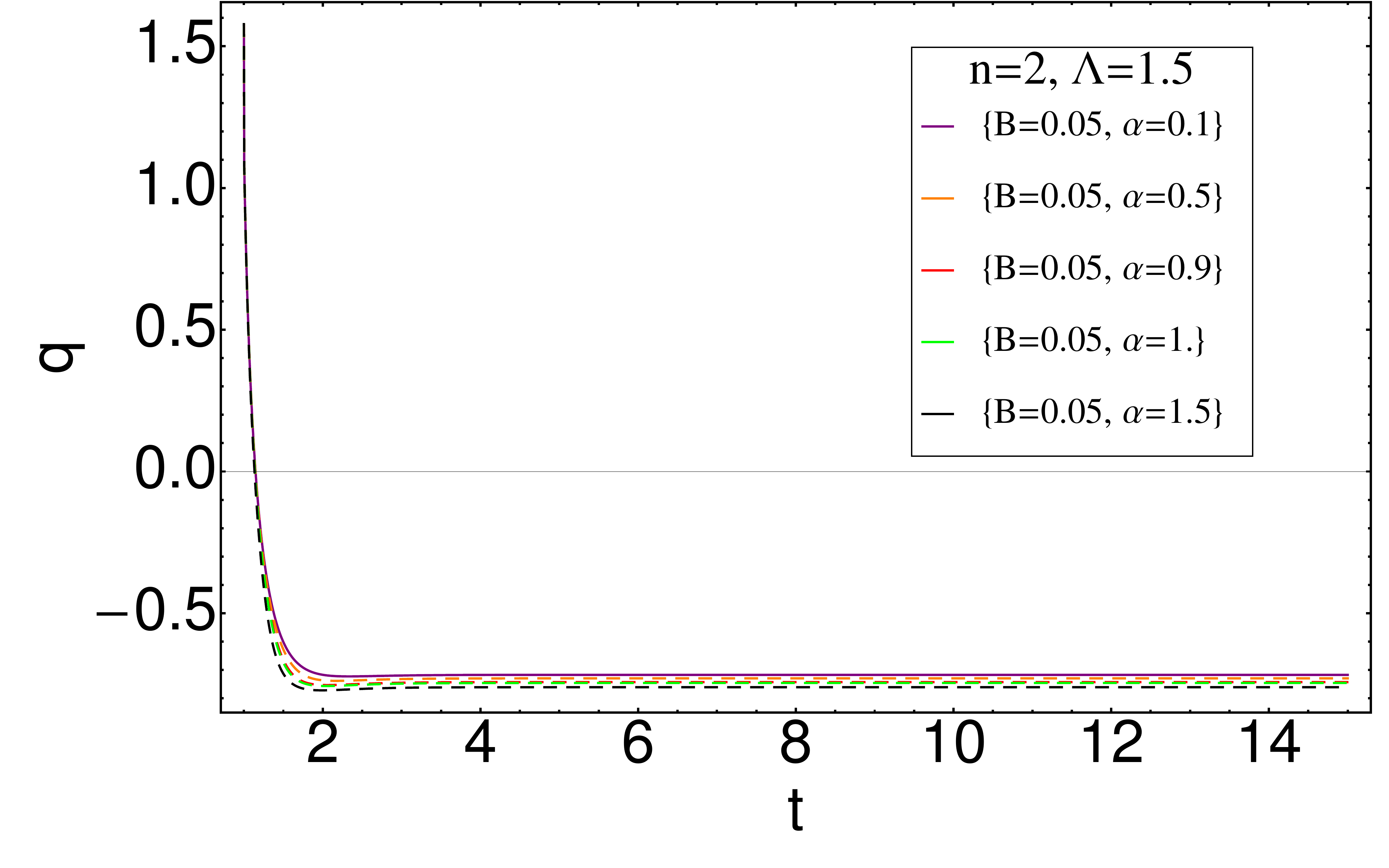} 
 \end{array}$
 \end{center}
\caption{The behavior of the Hubble parameter $H$ and the deceleration parameter $q$ against $t$ for the single fluid - extended Chaplygin gas Universe with constant $\Lambda$. The Hubble parameter defined as $H=\frac{\dot{a}(t)}{a(t)}$ and the deceleration parameter defined as $q = -1-\frac{\dot{H}}{H^{2}}$. }
 \label{fig:const1}
\end{figure}

\begin{figure}[h!]
 \begin{center}$
 \begin{array}{cccc}
\includegraphics[width=80 mm]{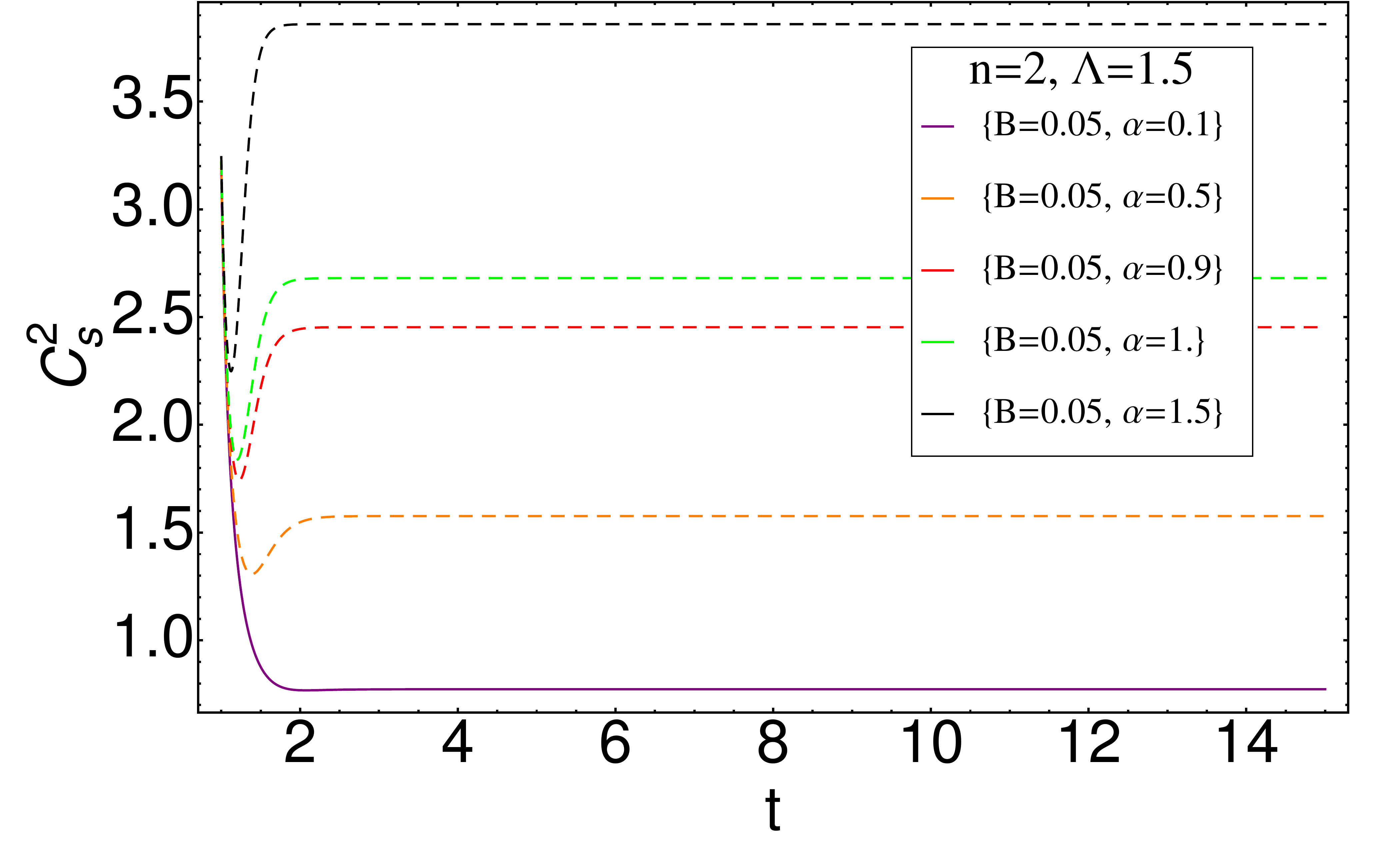} &
\includegraphics[width=80 mm]{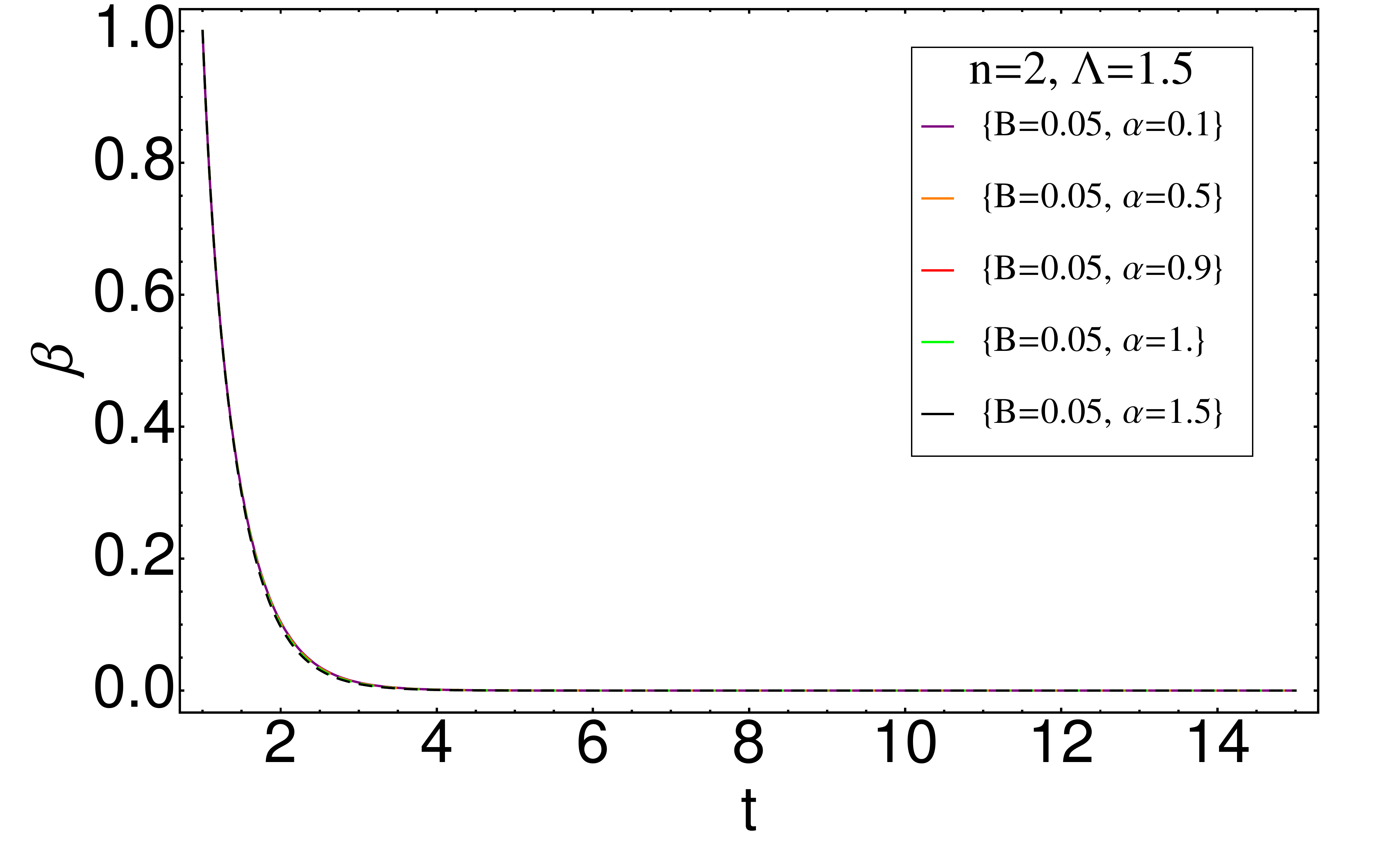}\\
 \end{array}$
 \end{center}
\caption{The behavior of the $C^{2}_{S}$ and $\beta$ against $t$ for the single fluid - extended Chaplygin gas Universe with constant $\Lambda$.}
 \label{fig:const2}
\end{figure}

The lack of a final fundamental physical theory to explain the dynamics of the Universe opens a great window for different speculations. Despite the huge effort still we are not able to answer to the fundamental open questions in a proper way which gives philosophical fights between different minds. To have accelerated expansion observed today and to have a possibility to explain this phenomenon in scope of GR we think about the existence of a "fluid" described by negative pressure and positive energy density, which wants to work against gravity. This fluid is known as dark energy(DE). In Literature we describe DE via EoS parameter, which determines the decay rate of the energy thus affecting the background expansion and the evolution of the matter perturbations,  however for the full description of the DE we need 2 more parameters. In a phenomenological approach we should use its sound speed $C_{s}$ and its anisotropic stress $\sigma$ \cite{Hu}. An accurate and deep understanding can provide $C_{s}$, which does not affect the background evolution, but a fundamental to characterize the perturbations. The anisotropic stress has been widely neglected and will be neglected in this work as well. The square of the sound speed is defined as
\begin{equation}
C^{2}_{S} = \frac{\delta P}{\delta \rho},
\end{equation}  
where $P$ is the pressure and $\rho$ is the energy density of the fluid. As a matter of fact, even a positive $C^{2}_{S}$ can rise a causality issue, therefore to have a correct physics we should have a well defined range. Surprisingly, by its simple form the $C^{2}_{S}$ is a good criterion to reject the theories~\cite {Hawking},~ \cite{Wald}. A widespread viewpoint is 
\begin{equation}\label{eq:causality}
0 \leq C^{2}_{S} \leq  1,
\end{equation}
which also can be challenged~\cite{Erickson},~ \cite{Ricardo}. Generally we could consider models with negative $C^{2}_{S}$ as well, however it is out from our general goal and we will not discuss our model from that perspective. We will also follow to the Eq.~(\ref{eq:causality}) to extract the part of the phenomenological model not to  violate the causality and to be physically meaningful. From the different observational data we know that we have a non static universe, while at the beginning Einstein believed in the static Universe, which assumed an existence of a negative pressure stopping the attraction of the universe, moreover we know that we have an accelerated expansion. According to the last observational data analysis we estimate the amount of DE to be $69\%$ of the Universe~\cite{Riess}~-~\cite{Verde}. The simple question of asking what is the nature of the DE is still one of the intriguing questions and left free space for various speculations, because existing theories and observational data are not able to give a final answer to this question. A big class of the DE models carries a phenomenological character and were introduced in cosmology by "hand". This is a working approach but not a satisfactory one, therefore the modifications of the field equations on the Lagrangian level could be considered more fundamental and satisfactory. This approach opened a wide range of the modified GR theories, but still with a big uncertainty, because the modifications have a crucial role in our understanding of the universe. In the modified theories a mathematical modification of the field equations gives rise to a term, which is eventually associated with the DE, and therefore DE becomes model dependent. Of course, we do not exclude the possibilities, that some of the approaches combined with particle physics, for example, could be useful and will illuminate phenomenology of the models completely. Consideration of the modified GR promises new insight into our understanding of the universe and it is one of the hot topics for the study. One of the goals of this work is to consider one of the modifications of GR and investigate the behavior of the universe in case of an extended Chaplygin gas~\cite{Kahya} with EoS given as
\begin{equation}\label{eq:ExtCh}
P_{Ch}=\sum_{k=1}^{n}{\frac{k^{2}}{1+k}\rho_{Ch}^{k}}-\frac{B}{\rho_{Ch}^{\alpha}},
\end{equation}
where $\alpha$ and $B$ are constants. We know that MCG is described by the EoS 
\begin{equation}
P = A\rho - \frac{B}{\rho^{\alpha}},
\end{equation}
The case of $A = 0$ recovers generalized Chaplygin gas EoS, and $A = 0$ together $\alpha = 1$ recovers the original Chaplygin gas EoS. The best fitted parameters are found to be $A = 0.085$ and $\alpha = 1.724$, while Constitution + CMB + BAO and Union + CMB + BAO results are $A = 0.061 \pm 0.079$, $\alpha = 0.053\pm0.089$, and $A = 0.110\pm0.097$, $\alpha = 0.089 \pm 0.099$ respectively~\cite{Xu},~\cite{Toribio}. Other observational constraints on modified Chaplygin gas model using Markov Chain Monte Carlo approach found that $A = 0.00189^{+0.00583}_{-0.00756}$, $\alpha = 0.1079^{+0.3397}_{-0.2539}$ at $1\sigma$ level and $A = 0.00189^{+0.00660}_{-0.00915}$ with $\alpha = 0.1079^{+0.4678}_{-0.2911}$ at $2\sigma$ level~\cite{Lu_Xu}.\\\\ 
An universe with the effective fluid described by Eq.~(\ref{eq:ExtCh}) in future discussing modified scenario with the constant $\Lambda$ shows that the accelerated expansion of the universe is possible to obtain. Moreover, we see that a transition from the decelerated phase to the accelerated one could take place during the evolution. The energy density of the extended Chaplygin gas will decrease through the evolution and eventually the universe will enter the phase with constant energy density of the gas. $\beta(t)$ parameter is also a decreasing function completely vanishing $t \to \infty$ which means that the dynamics of the universe for later stages of the evolution can be described by the field equations of GR with a constant $\Lambda$. With the appropriate values of the model parameters we can obtain a universe satisfying causality issue with the appropriate constraint on $C^{2}_{S}$ given by Eq.~(\ref{eq:causality}). Graphical behaviors of the the Hubble parameter $H$, the deceleration parameter $q$ with the behavior of $\beta(t)$ and $C^{2}_{S}$ can be found in Fig.-s~(\ref{fig:const1})~-~(\ref{fig:const2}) for
\begin{equation}
P_{Ch}=\frac{1}{2}\rho_{Ch}+\frac{4}{3}\rho^{2}_{Ch}-\frac{B}{\rho_{Ch}^{\alpha}},
\end{equation}
Which is due to the constraints on the model parameters from causality issue and $SneIa + BAO + CMB$ observational data  for distance modulus versus our theoretical results. We found the following constraints on the model parameters, which are in good agreement with the results obtained previously.\\
\begin{table}
  \centering
    \begin{tabular}{ | l | l | l | l | l | l | p{5cm} |}
    \hline
     $\alpha$ & $B$ & $n$ & $\Lambda$ & $H_{0}$ \\
  \hline
   $0\div 0.1$& $ 0.01 \div 0.08$ &$1\div 2$ & $0 \div 2$ & $0.84^{+0.1}_{-0.15}$ \\
    \hline
    \end{tabular}
\caption{ Values of the model parameters obtained from the $SneIa + BAO + CMB$ data  for distance modulus versus theoretical results for the single fluid - extended Chaplygin gas universe with  constant $\Lambda$. }
  \label{tab:myfirsttable}
\end{table}

Next,  we are interested in the other scenario where a coupling between the extended Chaplygin gas and DE also exists. In Literature readers could find a variety of models of the DE motivated from different physical theories. In this Paper we would like to work with one of them which is known as Quintessence DE with $\omega_{Q} > -1$~\cite{MKhurshudyan}~-~\cite{Saridakis}. Another class of quintessence DE models can include different modifications and extensions ~\cite{Peng}~-~\cite{Momeni}. The quintessence DE is a scalar field model described by a scalar field $\phi$ and self interacting potential $V(\phi)$. Energy density and pressure of such models are given as
\begin{equation}\label{eq:rhoQ}
\rho_{Q}=\frac{1}{2}\dot{\phi}^{2}+V(\phi),
\end{equation}   
and
\begin{equation}\label{eq:rhoP}
P_{Q}=\frac{1}{2}\dot{\phi}^{2}-V(\phi).
\end{equation}  
In our model the effective energy density and pressure assumed to be given as
\begin{equation}\label{eq:rhoeff}
\rho=\rho_{Q}+\rho_{Ch},
\end{equation}
and 
\begin{equation}\label{eq:Peff}
P=P_{Q}+P_{Ch},
\end{equation}
where $\rho_{Q}$ and $P_{Q}$ are energy density and pressure of the DE, while $P_{Ch}$ and $\rho_{Ch}$ represent pressure and energy density of the extended Chaplygin gas. The self interacting potential for the scalar field is taken to be of the exponential form  since exponential potentials are known to be significant in various cosmological models~\cite{Ferreira}~-~\cite{Gong}
\begin{equation}\label{eq:potV}
V(\phi)=V_{0}e^{ \left [-A \phi^{\gamma} \right ]},
\end{equation}
where $V_{0}$, $A$ and $\gamma$ are supposed to be constants. Moreover, related to the number of the field equations and unknowns (which will be discussed in the next section) we assume that the form of the effective varying $\Lambda(t)$ is also given and its dynamical form is 
\begin{equation}
\Lambda(t)=\rho_{Q}+\rho_{Ch}.
\end{equation}
Existing evidence of the coupling between DE and background matter is a hot discussion in theoretical Cosmology. From one side observations show the possibility of the coupling/interaction, but from the other side a fundamental theory answering why there should be interaction and how this coupling arose is missing. This makes good room for the phenomenological manipulations, but from the physical point of view it creates some uncertainties and is totally uncomfortable for works. There are many forms for the interaction term $Q$ (usual notation of the interaction) considered so far. On a phenomenological level we consider interactions based on  the dimensional analysis of the energy conservation equation giving hint that the dimension of $Q$ should be $[energy~density]/[time]$. In Ref.~\cite{Interaction_Saridakis} a new form for the term $Q$ was considered which for our model turns to be 
\begin{equation}\label{eq:Q2}
Q=H^{1-2m}b\rho_{Ch}^{m}\dot{\phi}^{2},
\end{equation}
with the Hubble parameter $H$, a constant $b$, energy density of the CG and field $\phi$. In the same work another interesting form for $Q$ also was considered
\begin{equation}\label{eq:Q3}
Q=H^{3-2m}b\rho^{m},
\end{equation}
where $b$ and $m$ are constants, which with $m=1$ will reduce to $Q=Hb\rho$: the form intensively considered in Literature.\\\\
The paper is organized as follow: In the "Introduction" section, besides the introduction, the general behavior of the single fluid cosmological model with the constant $\Lambda$ in Lyra geometry is considered. In section "The field equations" we discuss modified field equations and introduce interaction term $Q$ into dynamics of the universe. In section "Model and physics" we discuss behavior of the model resulting from the causality issue and observational constraints. In the last section we give conclusion.

\section{\large{The field equations}}
Lyra geometry is an example of scalar tensor theory and one of the modifications of GR suggested by Lyra  as a modification of Riemannian  geometry~\cite{Lyra}. In this modification the Weyl's gauge is modified. Field equations constructed an analogue of the Einstein field equations based on Lyra's geometry that can be written as~\cite{Sen},~\cite{Dunn}
\begin{equation}
R_{\mu\nu}-\frac{1}{2}g_{\mu\nu}R+\frac{3}{2}\psi_{\mu}\psi_{\nu}-\frac{3}{4}g_{\mu \nu}\psi^{\alpha}\psi_{\alpha}=T_{\mu\nu}.
\end{equation}
It was pointed out that the constant displacement field $\psi_{\alpha}$ of this theory can be interpreted as a cosmological constant $\Lambda$ in the normal relativistic treatment \cite{Halford}. We are interested in the other modification of the field equations which contain varying cosmological constant $\Lambda(t)$ and which can be written as~\cite{Shchigolev},~\cite{Martiros}~-~\cite{Martiros3}
\begin{equation}\label{eq:Einstein eq}
R_{\mu\nu}-\frac{1}{2}g_{\mu\nu}R-\Lambda g_{\mu \nu}+\frac{3}{2}\psi_{\mu}\psi_{\nu}-\frac{3}{4}g_{\mu \nu}\psi^{\alpha}\psi_{\alpha}=T_{\mu\nu}.
\end{equation}
Considering the content of the Universe to be a perfect fluid, we will have
\begin{equation}\label{eq:T}
T_{\mu\nu}=(\rho+P)u_{\mu}u_{\nu}-Pg_{\mu \nu},
\end{equation}
where $u_{\mu}=(1,0,0,0)$ is a 4-velocity of the co-moving
observer, satisfying $u_{\mu}u^{\mu}=1$. Let $\psi_{\mu}$ be a time-like
vector field of displacement
\begin{equation}
\psi_{\mu}=\left ( \frac{2}{\sqrt{3}}\beta(t),0,0,0 \right ),
\end{equation}
where $\beta=\beta(t)$ is a function of time alone, and the factor $\frac{2}{\sqrt{3}}$ is substituted in order to simplify the writing of all the following equations. By using FRW metric for a flat universe
\begin{equation}\label{s2}
ds^2=-dt^2+a(t)^2\left(dr^{2}+r^{2}d\Omega^{2}\right),
\end{equation}
field equations can be reduced to the following Friedmann equations
\begin{equation}\label{eq:Freidmann1}
3H^{2}-\beta^{2}=\rho+\Lambda,
\end{equation}
and
\begin{equation}\label{eq:Freidmann2}
2\dot{H}+3H^{2}+\beta^{2}=-P+\Lambda,
\end{equation}
where $H=\frac{\dot{a}}{a}$ is the Hubble parameter, and an overdot stands for differentiation with respect to cosmic time $t$, $d\Omega^{2}=d\theta^{2}+\sin^{2}\theta d\phi^{2}$, and $a(t)$
represents the scale factor. The $\theta$ and $\phi$ parameters are the usual azimuthal and polar angles of spherical coordinates, with $0\leq\theta\leq\pi$ and $0\leq\phi<2\pi$. The coordinates ($t, r,
\theta, \phi$) are called co-moving coordinates. Cosmological models with reach and interesting physics are possible to build using metrics different than $FRW$. It is an interesting research to consider cosmological models with anisotropic DE~\cite{Samanta}, because it was demonstrated that we can construct models for the Universe approaching isotropy even in the presence of an anisotropic fluid, moreover fluid also isotropizes in the accelerated expanding Universe. As such models are not the object of our resent investigation we found appropriate to refer our readers to~\cite{Samanta} and references therein for more information on such models.  \\ \\
From Eq.-s~(\ref{eq:Freidmann1})~-~(\ref{eq:Freidmann2}) directly follows that the continuity equation reads as
\begin{equation}\label{eq:coneq}
\dot{\rho}+\dot{\Lambda}+2\beta\dot{\beta}+3H(\rho+P+2\beta^{2})=0.
\end{equation}
With 
\begin{equation}\label{eq:DEDM}
\dot{\rho}+3H(\rho+P)=0,
\end{equation}
Eq.~(\ref{eq:coneq}) will give a link between $\Lambda(t)$ and $\beta(t)$ of the following form
\begin{equation}\label{eq:lbeta}
\dot{\Lambda}+2\beta\dot{\beta}+6H\beta^{2}=0.
\end{equation}
To introduce an interaction between DE and DM, usually, we mathematically split  Eq.~(\ref{eq:DEDM})  into two  equations where interaction term $Q$ enters explicitly and controls the dynamics of the energy densities of the components  as
\begin{equation}\label{eq:inteqm}
\dot{\rho}_{DM}+3H(\rho_{DM}+P_{DM})=Q,
\end{equation}
and
\begin{equation}\label{eq:inteqG}
\dot{\rho}_{DE}+3H(\rho_{DE}+P_{DE})=-Q.
\end{equation}
We start analysis of the two-fluid component cosmological model in the next section.
\section{\large{The Model and physics}}
\begin{figure}[h!]
 \begin{center}$
 \begin{array}{cccc}
\includegraphics[width=80 mm]{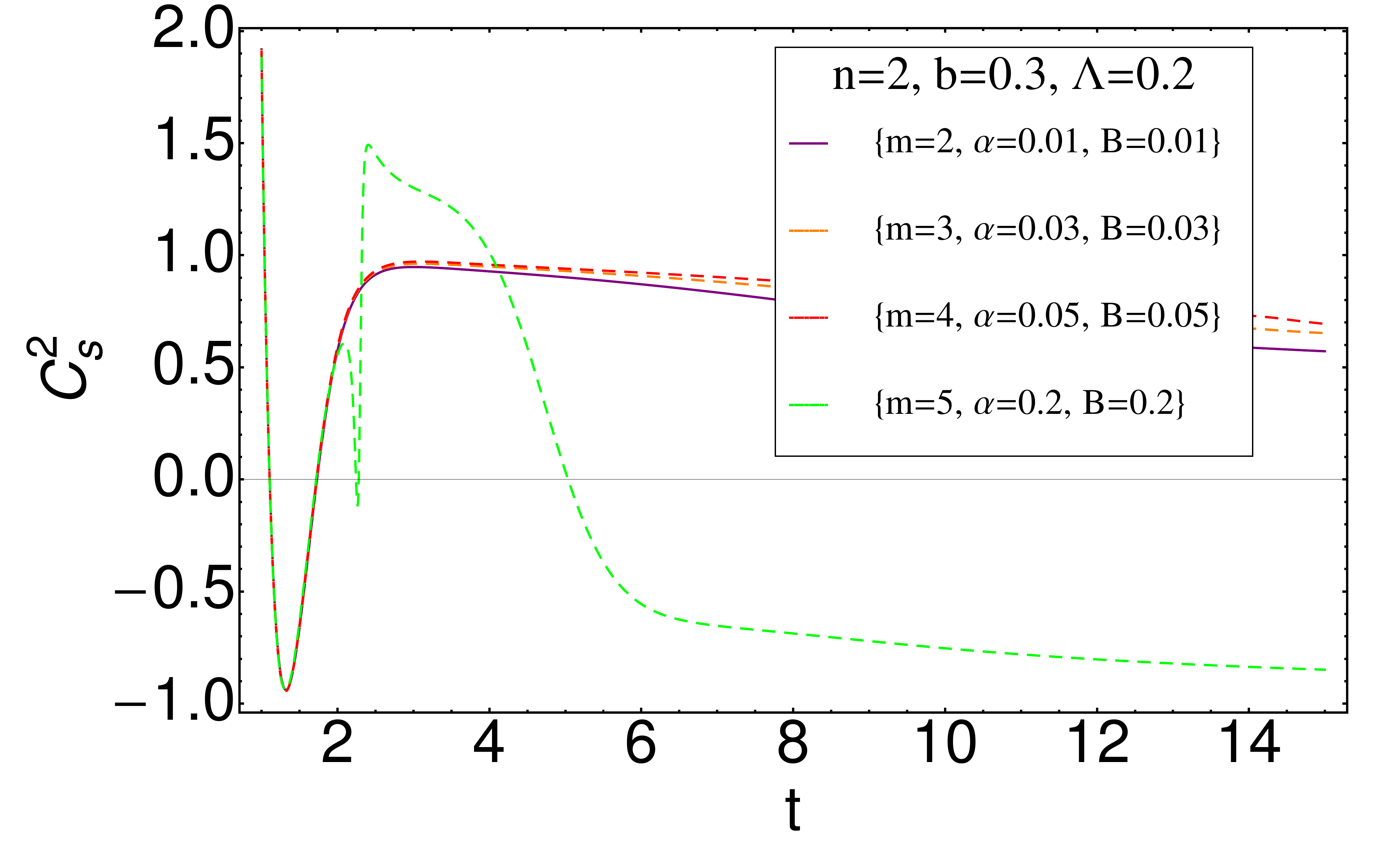} &
\includegraphics[width=80 mm]{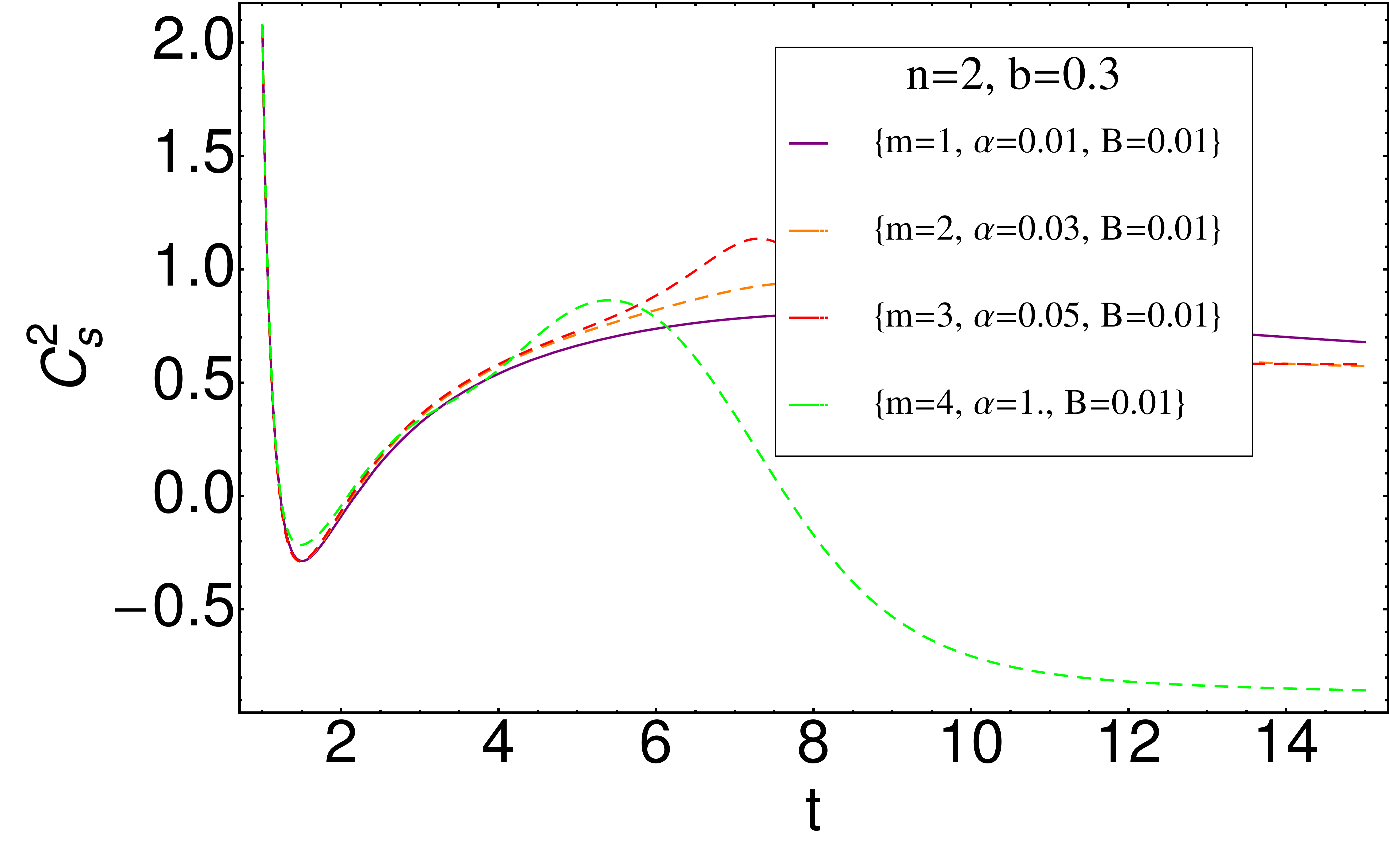}\\
 \end{array}$
 \end{center}
\caption{Behavior of the $C^{2}_{S}$ against $t$ for the effective two-component cosmic fluid for constant $G$ and $\Lambda$ represents the left plot. The right plot represents the case corresponding to the $\Lambda(t)=\rho_{Ch}+\rho_{Q}$ with constant $G$.}
 \label{fig:modelC}
\end{figure}

\begin{figure}[h!]
 \begin{center}$
 \begin{array}{cccc}
\includegraphics[width=80 mm]{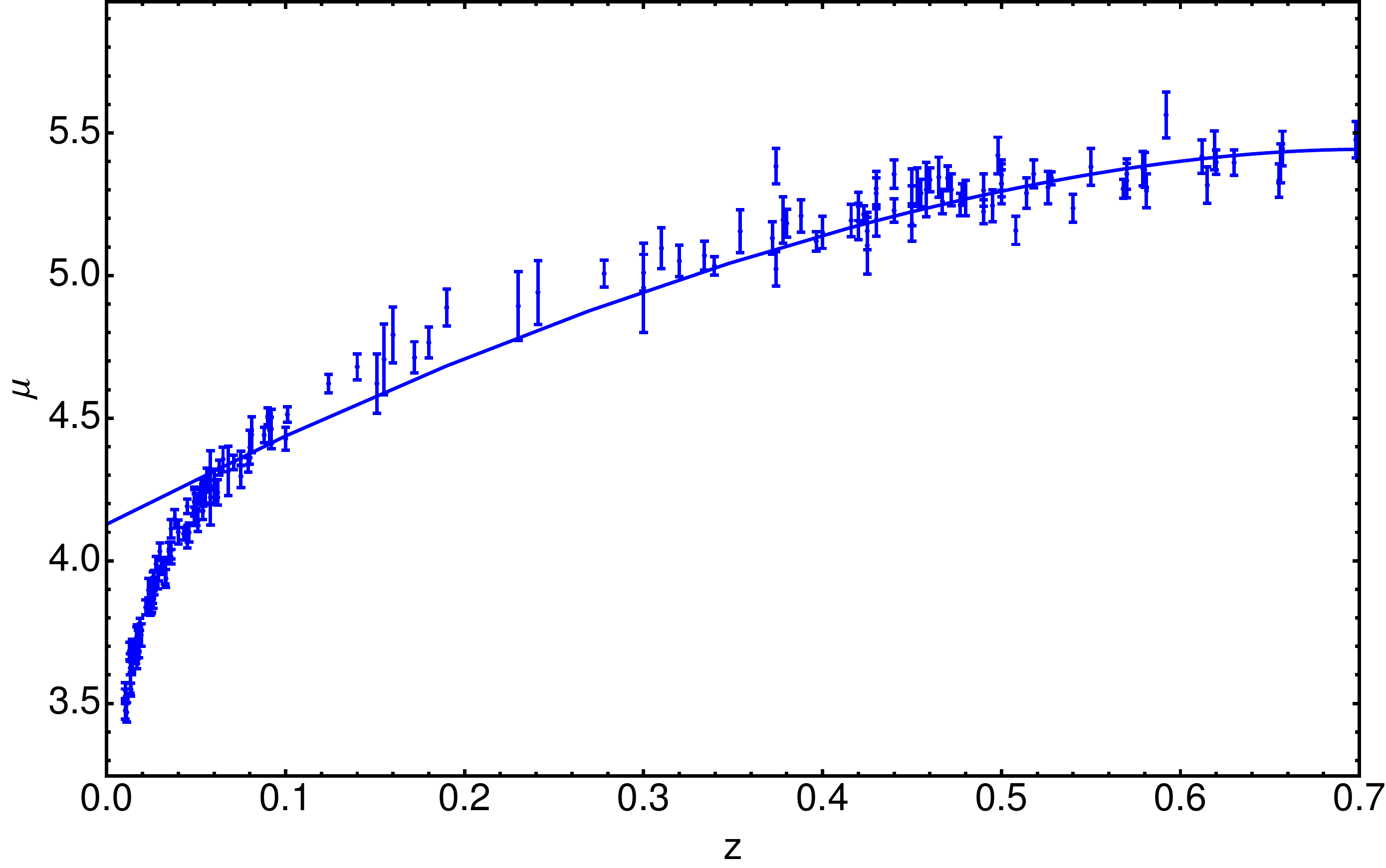} &
\includegraphics[width=80 mm]{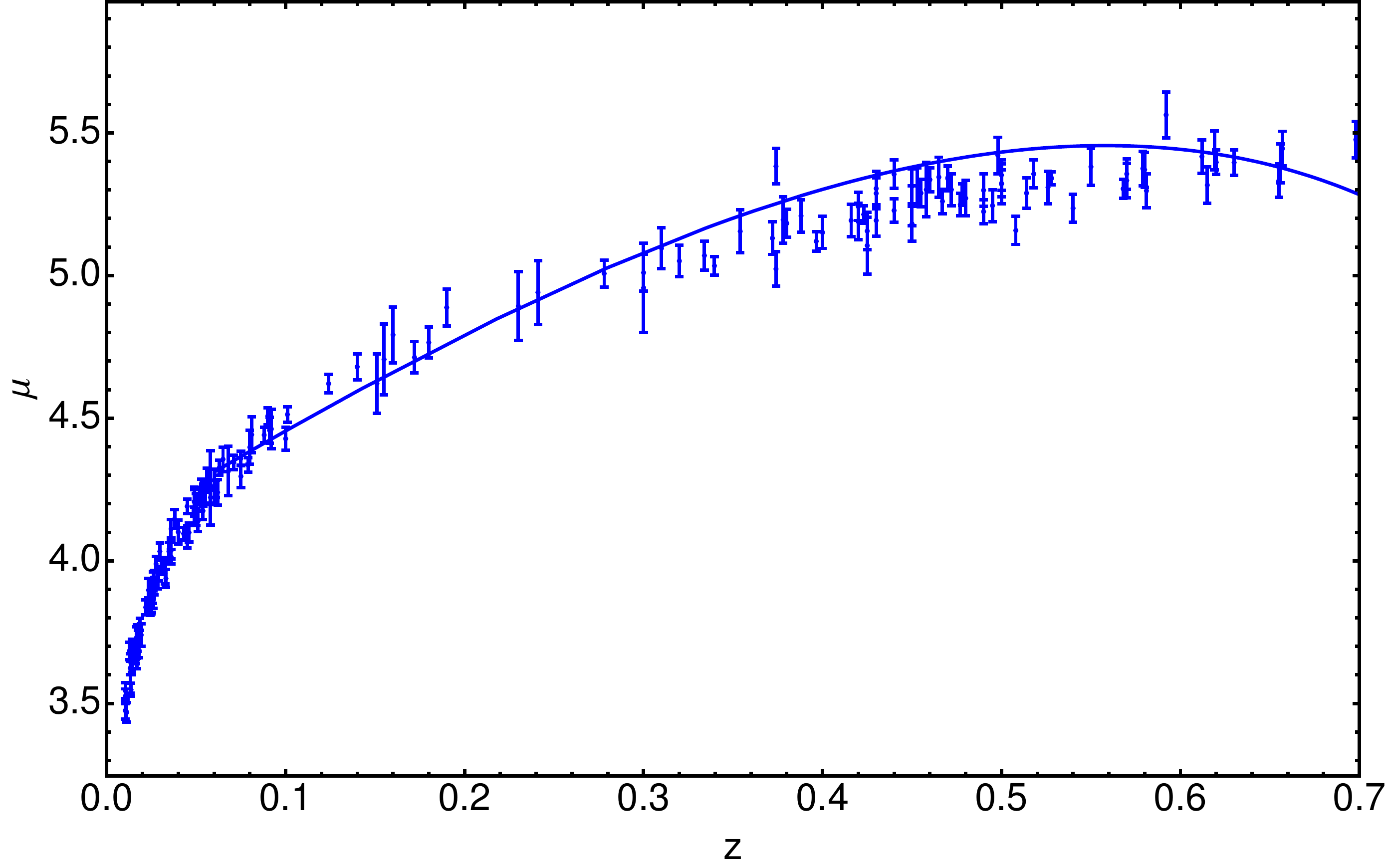}\\
 \end{array}$
 \end{center}
\caption{Observational data SneIa + BAO + CMB for distance modulus versus our theoretical results. Left plot corresponds to the constant $\Lambda$ for the universe with two-component effective fluid. Right plot with the same setup for $\Lambda(t)=\rho_{Ch}+\rho_{Q}$.}
 \label{fig:modelobs}
\end{figure}

\begin{figure}[h!]
 \begin{center}$
 \begin{array}{cccc}
\includegraphics[width=80 mm]{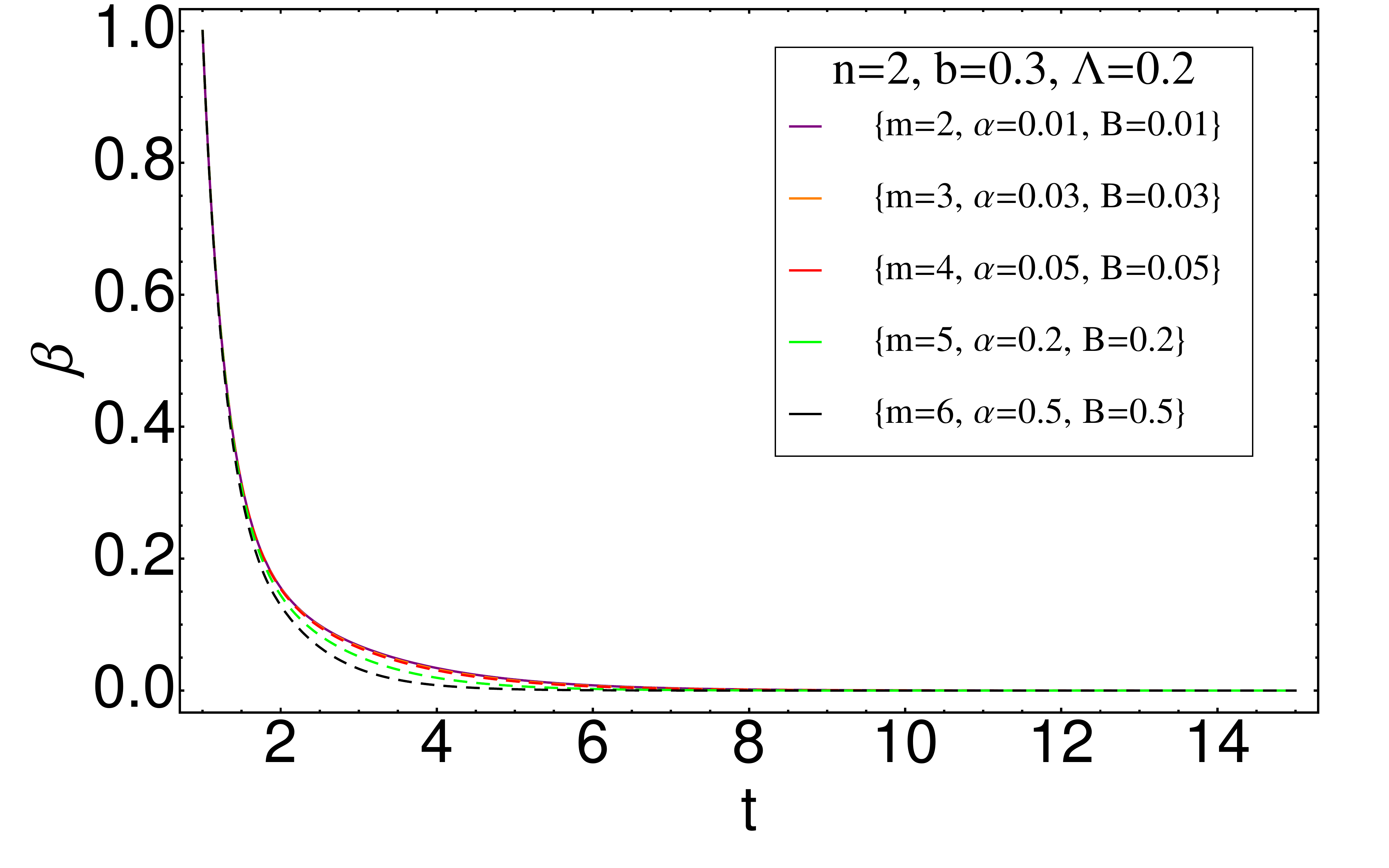} &
\includegraphics[width=80 mm]{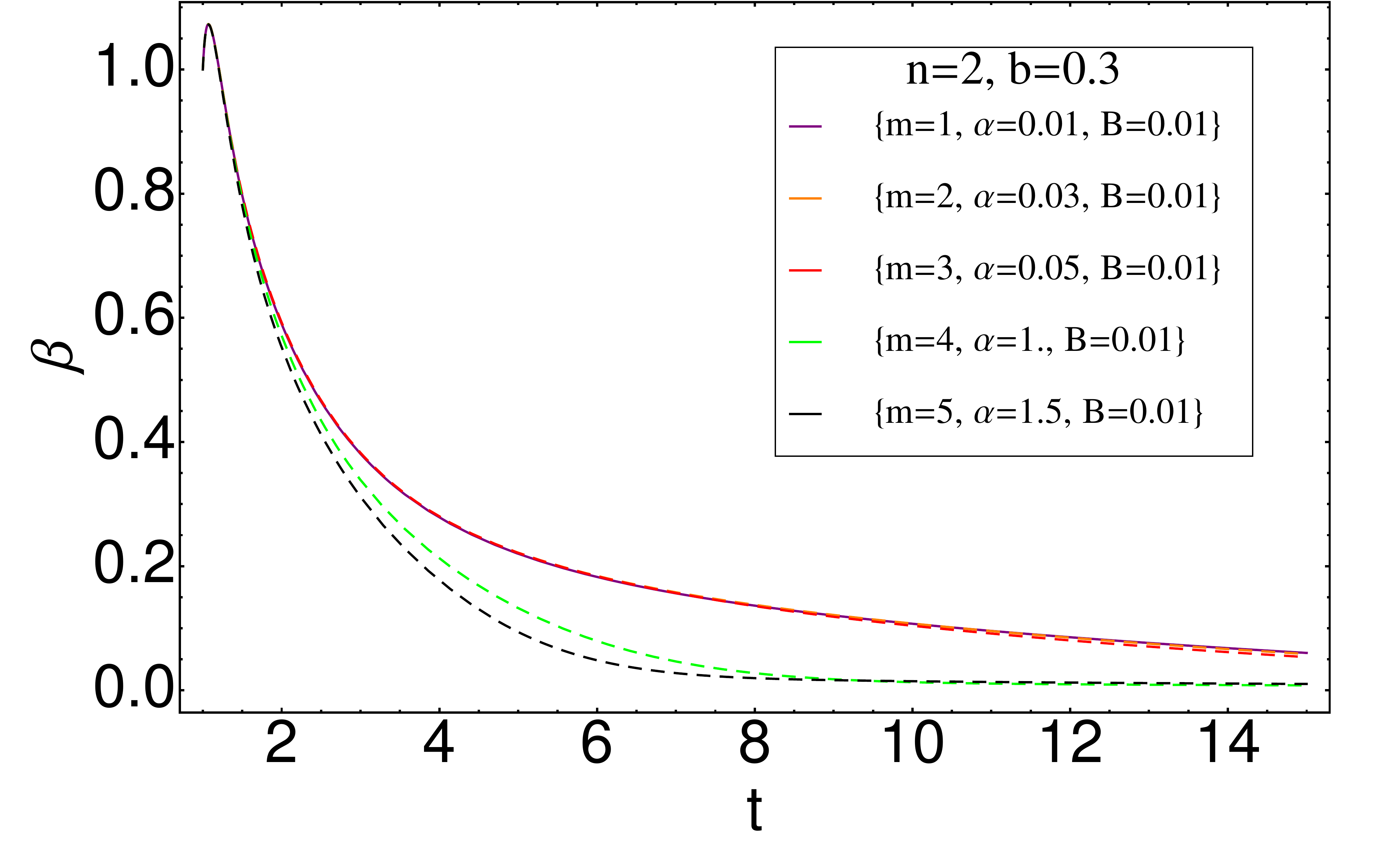}\\
 \end{array}$
 \end{center}
\caption{Behavior of $\beta(t)$ against $t$ for the effective two-component cosmic fluid for constant $G$ and $\Lambda$ is represented in the left plot. The right plot represents the case which corresponds to the $\Lambda(t)=\rho_{Ch}+\rho_{Q}$ with the constant $G$.}
 \label{fig:modelbeta}
\end{figure}

\begin{figure}[h!]
 \begin{center}$
 \begin{array}{cccc}
\includegraphics[width=80 mm]{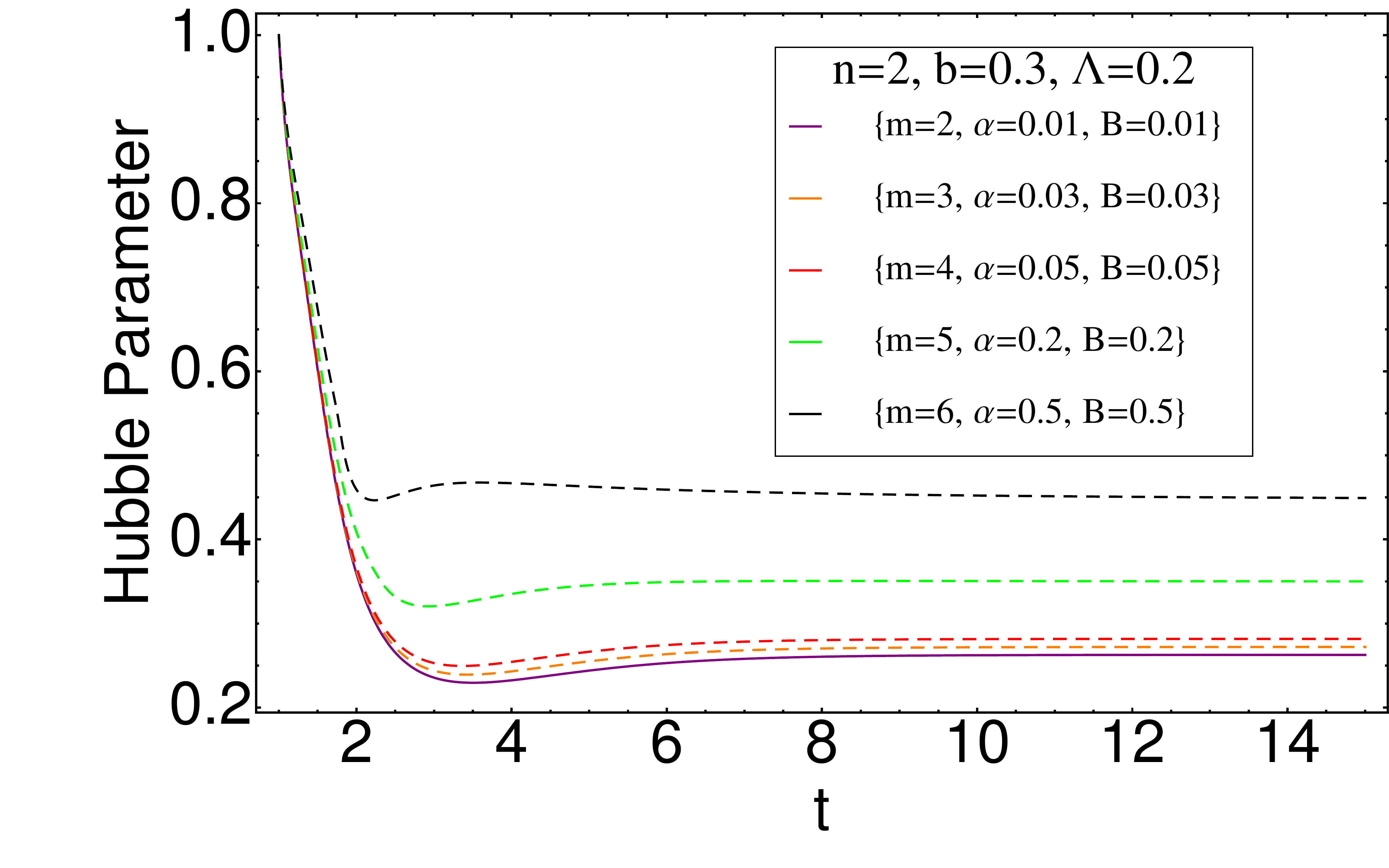} &
\includegraphics[width=80 mm]{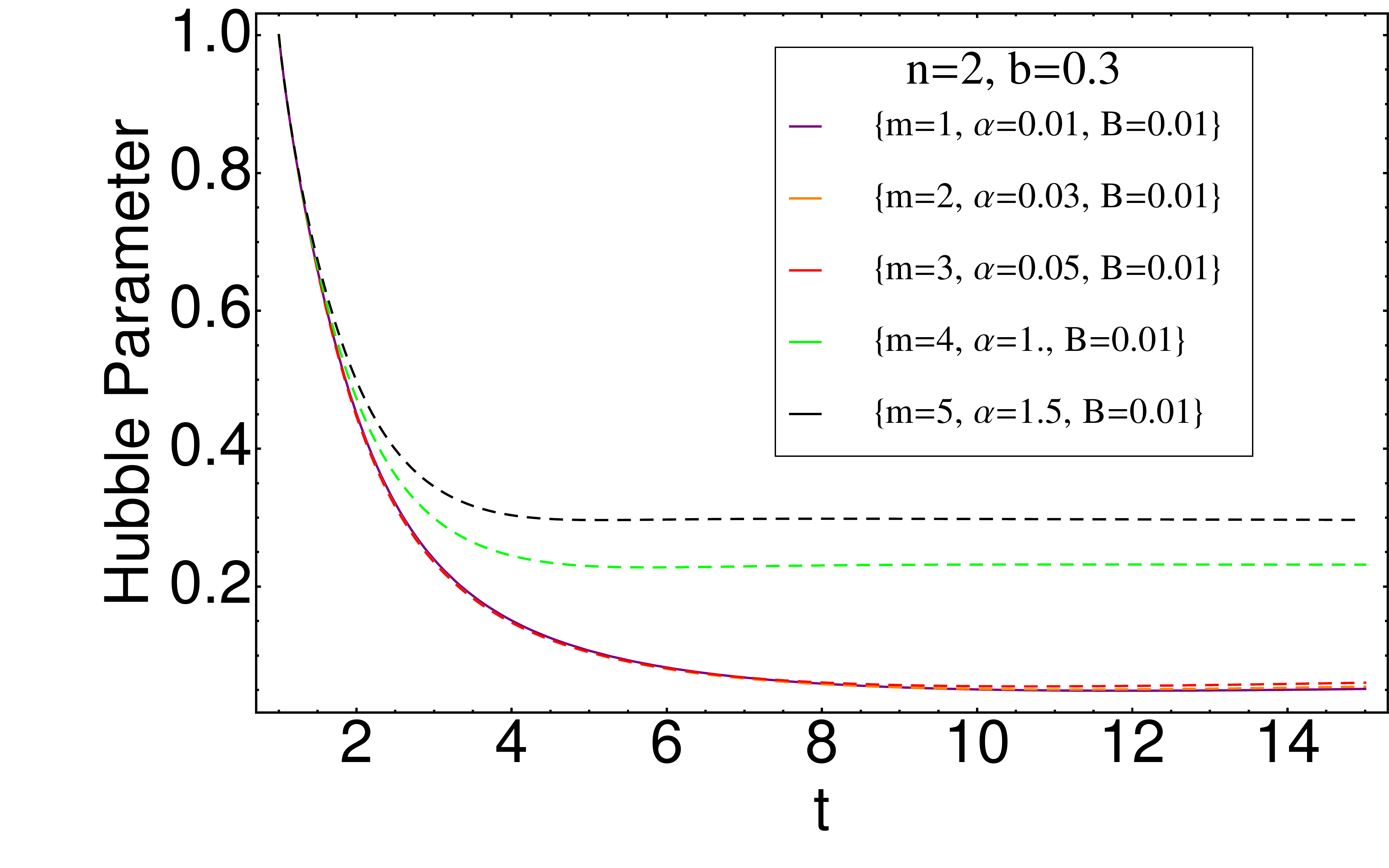}\\
 \end{array}$
 \end{center}
\caption{Behavior of the Hubble parameter $H$ against $t$ for the constant $G$ and $\Lambda$ is represented in the left plot. The right plot represents the universe with $\Lambda(t)=\rho_{Ch}+\rho_{Q}$  and constant $G$ . The Hubble parameter defined as $H=\frac{\dot{a}(t)}{a(t)}$. }
 \label{fig:modelH}
\end{figure}

\begin{figure}[h!]
 \begin{center}$
 \begin{array}{cccc}
\includegraphics[width=80 mm]{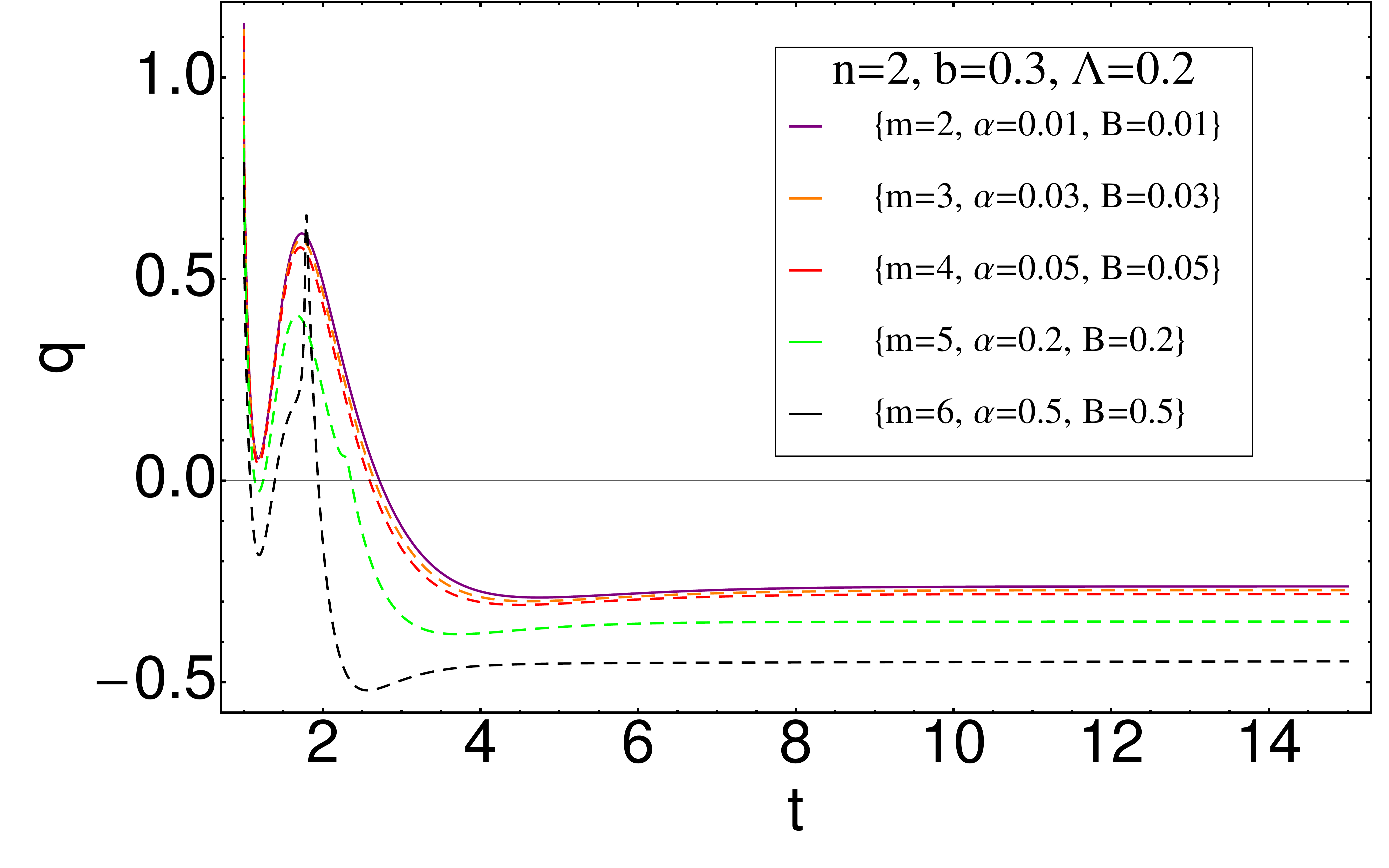} &
\includegraphics[width=80 mm]{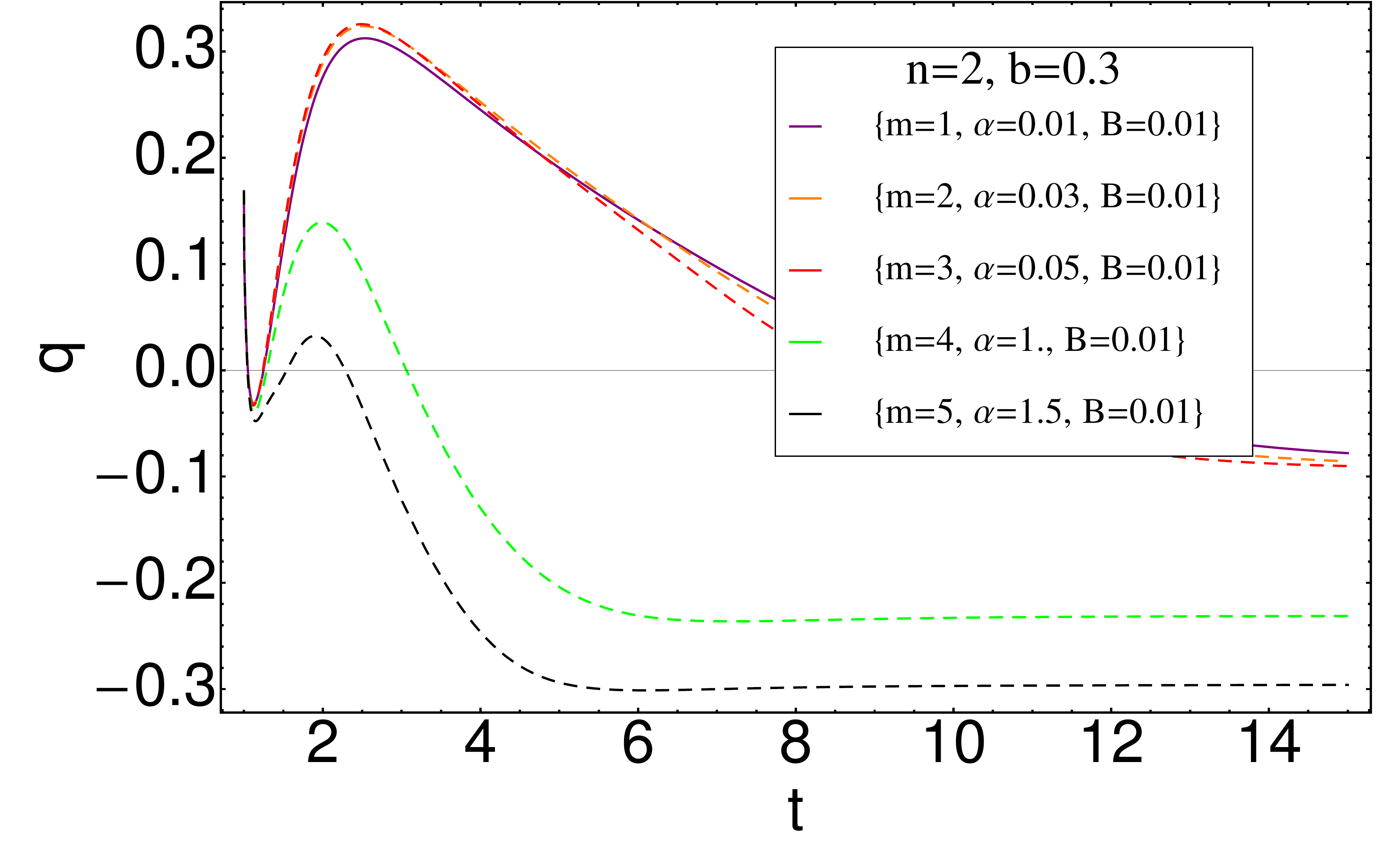}\\
 \end{array}$
 \end{center}
\caption{Behavior of the deceleration parameter $q$ against $t$ for the constant $G$ and $\Lambda$ is represented in the left plot. The right plot represents the universe with $\Lambda(t)=\rho_{Ch}+\rho_{Q}$ and constant $G$. The deceleration parameter defined as $q = -1-\frac{\dot{H}}{H^{2}}$.}
 \label{fig:modelq}
\end{figure}

\begin{figure}[h!]
 \begin{center}$
 \begin{array}{cccc}
\includegraphics[width=80 mm]{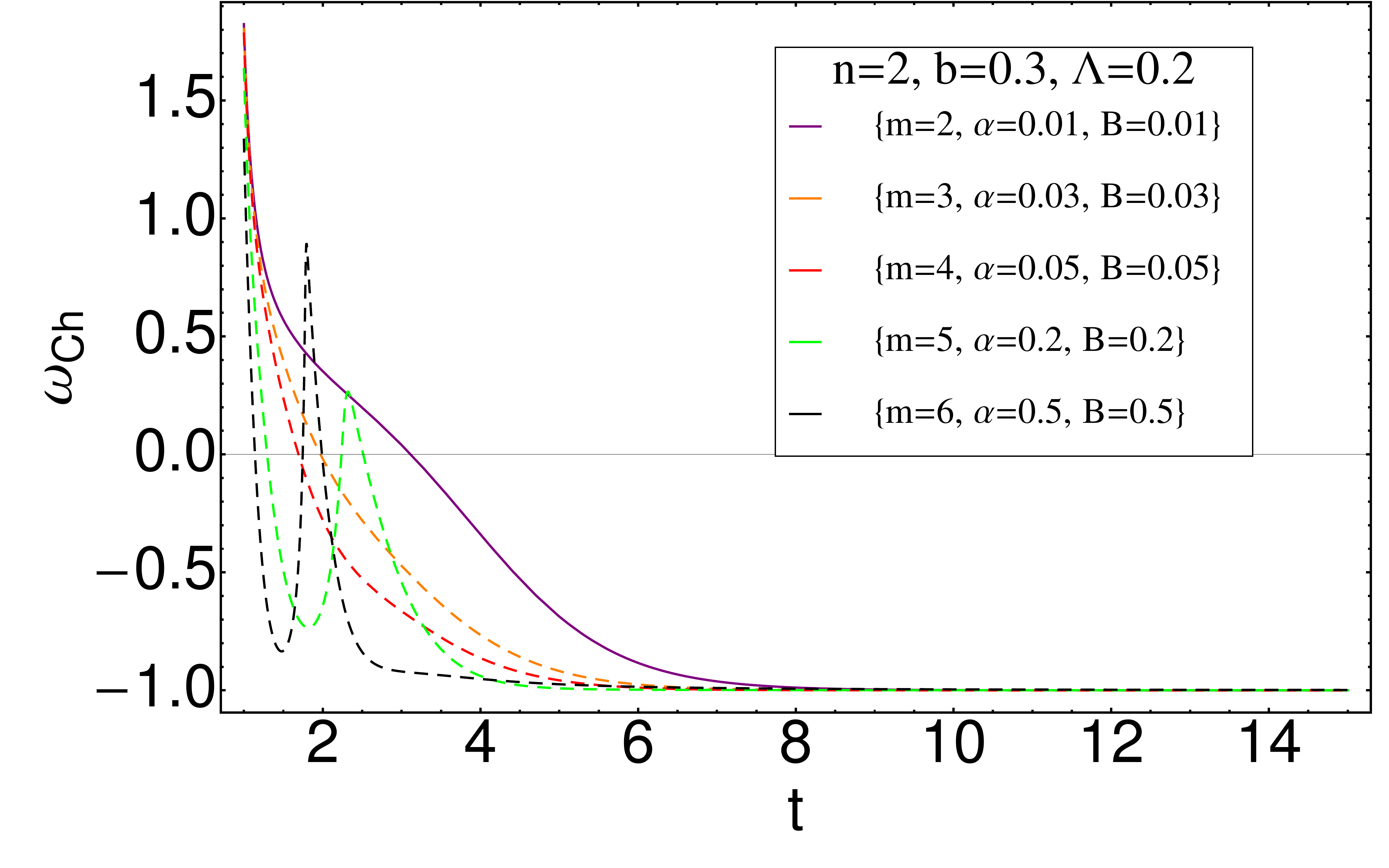} &
\includegraphics[width=80 mm]{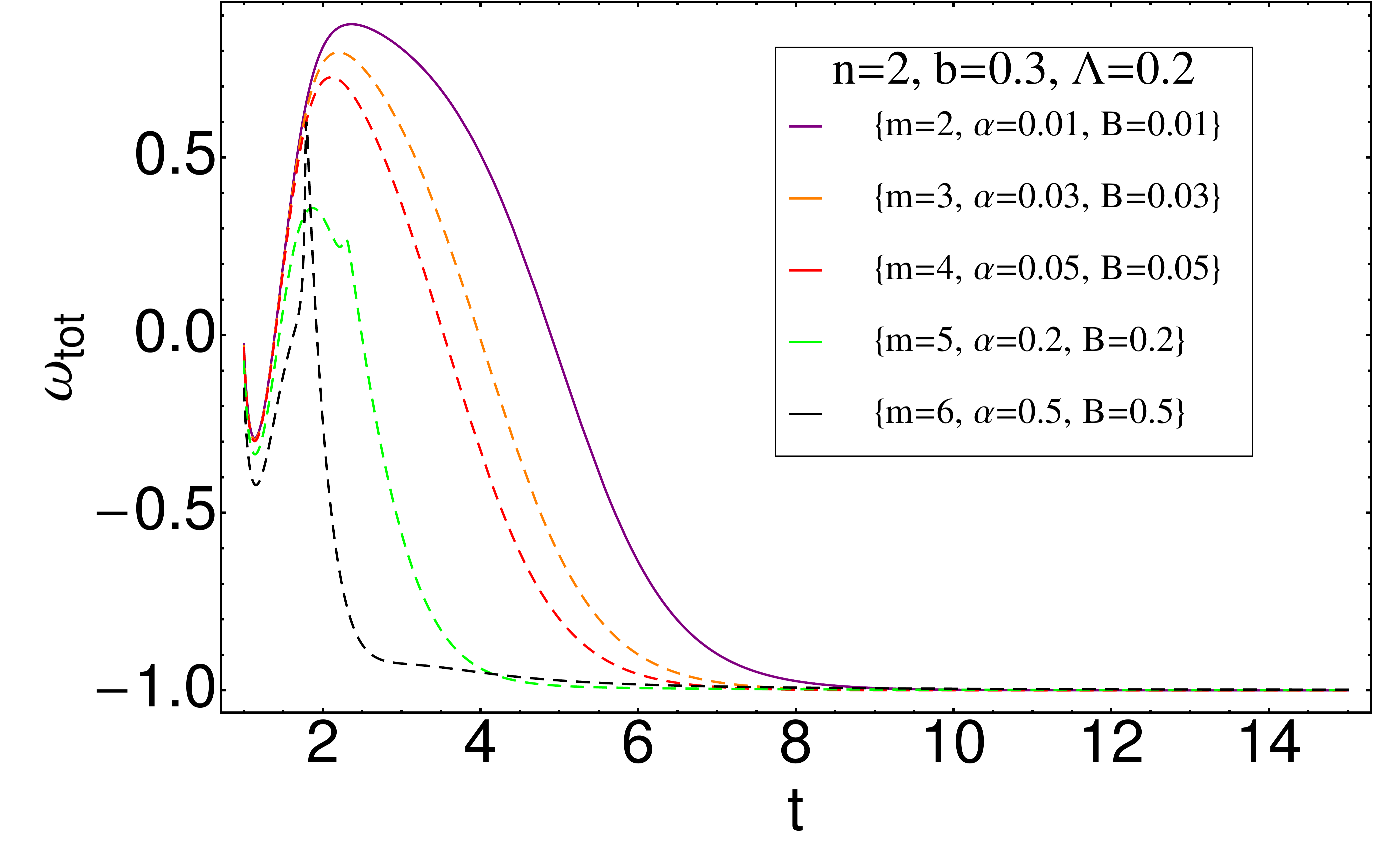} \\
\includegraphics[width=80 mm]{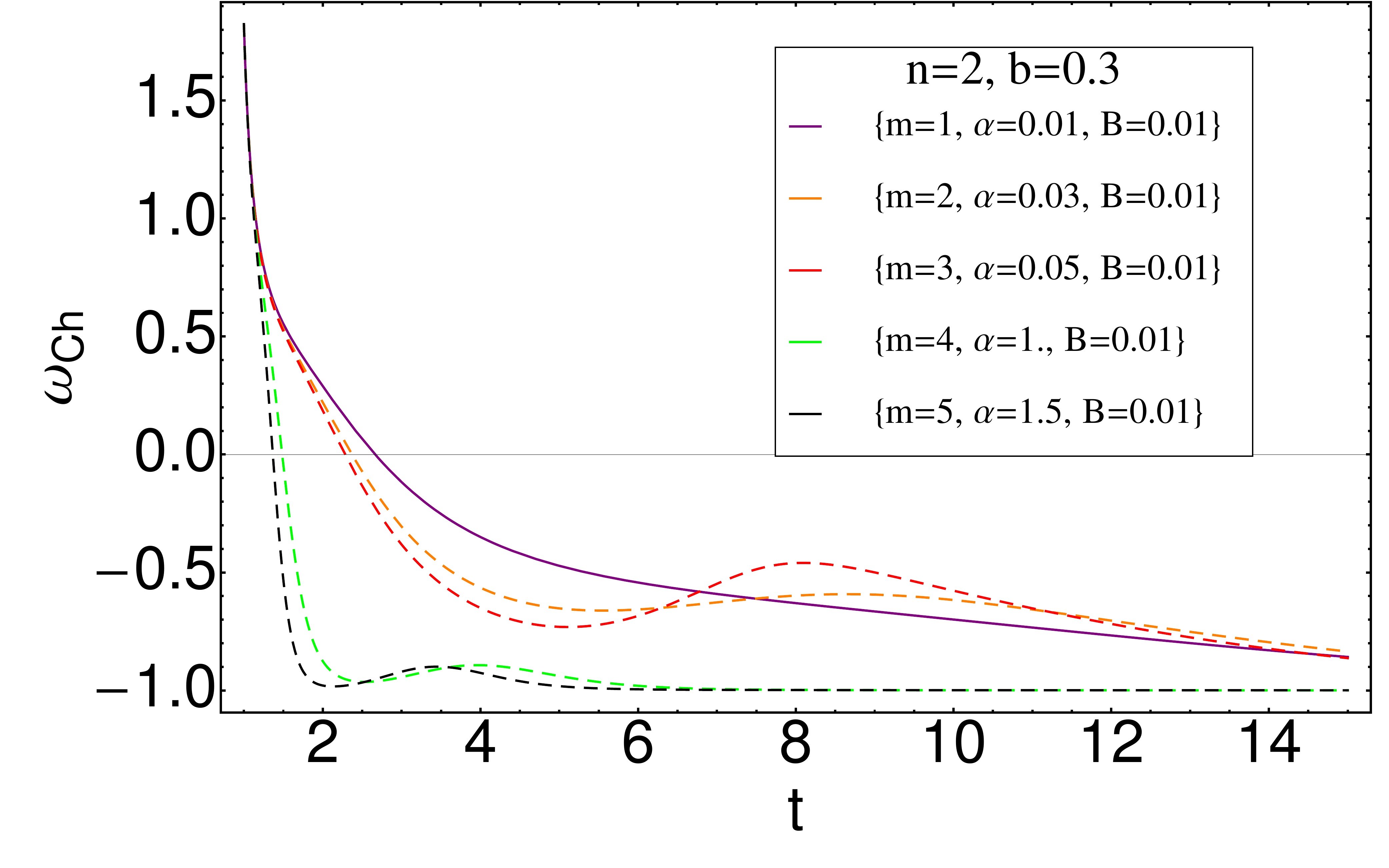}&
\includegraphics[width=80 mm]{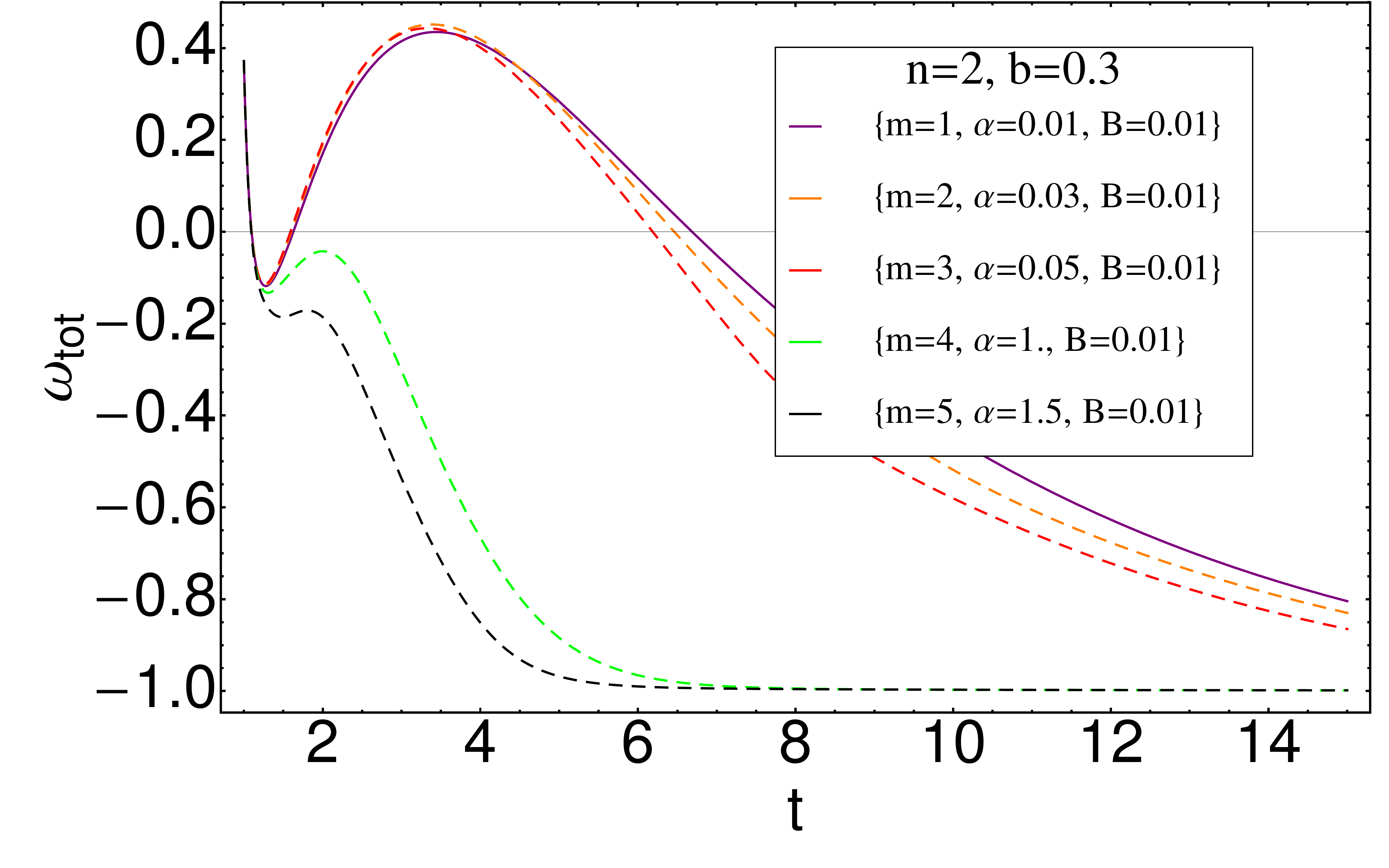} \\
 \end{array}$
 \end{center}
\caption{Behavior of the EoS parameters $\omega_{Ch}$ and $\omega_{tot}$ against $t$ for the constant $G$ and $\Lambda$ represents the top panels. The bottom panel with two plots represents the behavior of the EoS parameters $\omega_{Ch}$ and $\omega_{tot}$ assuming $\Lambda(t)=\rho_{Ch}+\rho_{Q}$ with the constant $G$. The EoS parameters for the gas and the two-component fluid read as $\omega_{Ch}=\frac{ P_{Ch} } {\ rho_{Ch} }$ and $\frac{ P_{Ch}+P_{Q} } { \rho_{Ch}+\rho_{Q} }$.}
 \label{fig:modelomega}
\end{figure}

We would like to start our discussion with the model where we have constant gravitational $G$ and cosmological $\Lambda$ constants. Such consideration simplifies the field equations under consideration. 
According to this assumption we can see that Eq.~(\ref{eq:coneq}) will be modified to
\begin{equation}
\dot{\rho}+2\beta\dot{\beta}+3H(\rho+P+2\beta^{2})=0,
\end{equation}
and with the other assumption $\dot{\rho}+3H(\rho+P)=0$ we will have a differential equation for the dynamics of the $\beta(t)$
\begin{equation}
\dot{\beta}+3H\beta=0.
\end{equation}
An integration of the last equation reveals a dependence between $\beta(t)$ and the scale factor $a(t)$ as following
\begin{equation}
\beta=\beta_{0}a^{-3},
\end{equation} 
where $\beta_{0}$ is the integration constant. The explicit form for the Hubble parameter can be used here which can be obtained from Eq.~(\ref{eq:Freidmann1}) 
\begin{equation}
H=\sqrt{\frac{\rho+\Lambda+a_{0}a^{-6}}{3}}.
\end{equation}
We see that $\beta(t) \to 0$ when $a \to \infty$. The interaction $Q=H^{1-2m}b\rho_{Ch}^{m}\dot{\phi}^{2}$ will give us the following two differential equations describing the dynamics of the energy densities of the fluid components
\begin{equation}
\ddot{\phi}+3H\dot{\phi}+H^{1-2m}b\rho_{Ch}^{m}\dot{\phi}+\frac{dV(\phi)}{d\phi}=0,
\end{equation}
and
\begin{equation}
\dot{\rho}_{Ch}+3H(1+\omega_{Ch})\rho_{Ch}-H^{1-2m}b\dot{\phi}^{2}\rho_{Ch}^{m}=0,
\end{equation}
where $b$ and $m$ are constants. Further analysis of the model assuming a dynamical form for the $\Lambda(t)$ is performed numerically and the results are discussed below in a proper way with great accuracy. We also would like to mention that our main goal is to illuminate a realistic part of our phenomenological model using physics related to the causality issue, which directly imposes constraints on the square of the sound speed $C^{2}_{S}$ to be a positive and between $0$ and $1$. In both cases: constant $\Lambda$ and $\Lambda(t)$, we fix values of the parameters in such a way as to have a stable cosmological effective two-component fluid and DE at least for the later stages of the evolution (Fig.~\ref{fig:modelC}).
Moreover, we apply $SneIa + BAO + CMB$ observational constraints on the model to find the best fit of the theoretical model with observations (Fig.~(\ref{fig:modelobs})), from the other hand this will narrow the possible range of the values of the model parameters and we will have an appropriate model.
In case of the constant $\Lambda$ we found $\beta(t) \to 0$, which was and seen in the case of the single fluid model presented in "Introduction" section and we have the same conclusion about the theory describing the old universe. Cosmology corresponding to the varying $\Lambda$ shows a different behavior and can not be described by the field equations of GR, because $\beta(t)$ is not a vanishing function anymore and admits a positive small constant value (Fig.~\ref{fig:modelbeta}).  
For our model we see that the Hubble parameter (presented in Fig~(\ref{fig:modelH})) in both cases is a decreasing function and $H \to const$ when $t \to \infty$. It is positive from the beginning of the cosmic evolution to the end, therefore we will have only expanding universe. The deceleration parameter for the universe with constant $\Lambda$ and $G$ is a positive at the beginning, but we observe that in this scenario transition to $q<0$ could be seen in the early history of the universe, which could be changed again with transition to $q>0$. However our universe will enter the old universe phase with constant and negative deceleration parameter $q$. When we consider a dynamical $\Lambda(t)$, basically we see the same behavior compared to the previous case with one difference, that the transition amplitude and time scales can be changed (to be longer). The discussed behavior of the deceleration parameter $q$ can be found in left and right plots of the Fig.~(\ref{fig:modelq}). At this step we can conclude that with our model we can have a universe with accelerated expansion. Furthermore, the next important step will be understanding the behavior of the content of the Universe, therefore we would like to discuss the behavior of the EoS parameters of the fluids and the effective interacting fluid. We saw that our universe can be in several different phases, but eventually will enter de-Sitter Universe with $\omega_{tot} = -1$ (Fig.~(\ref{fig:modelomega})). We observe also interesting oscillatory behavior in $\omega_{Ch}$ in case of the varying  $\Lambda(t)$ for the relatively early stages of the evolution which vanished completely in the old universe. Our universe for latter stages of the evolution will be only in the quintessence phase.  Further analysis revealed that we have interesting constraints on the model parameters and they are in good cooperation with the results from the Literature obtained before. For instance, in the universe  with the constant $\Lambda$ we obtained $0.01 \leq \alpha \leq 0.05$ and $0.01\leq \alpha \leq 0.07$ for the universe in case of the varying $\Lambda$.  Further information for the other parameters can be found in Table~\ref{tab:2Table}. Graphical behaviors of the cosmological parameters for the two-component fluid universe corresponds to the extended Chaplygin gas of the form
\begin{equation}
P_{Ch}=\frac{1}{2}\rho_{Ch}+\frac{4}{3}\rho^{2}_{Ch}-\frac{B}{\rho_{Ch}^{\alpha}},
\end{equation}
which provides the best fit of the cosmological model with the observational data.

\begin{table}
  \centering
    \begin{tabular}{ | l | l | l | l | l | l | l | p{5cm} |}
    \hline
    $Model$ & $\alpha$ & $B$ & $n$ & $\Lambda$ & $m$ \\
  \hline
   $\Lambda = const$ & $0.01\div 0.05$& $0.01 \div 0.05$ &$1\div 2$ & $0\div 0.3$ & $0\div 3$ \\
    \hline
   $\Lambda(t) = \rho_{Q}+\rho_{Ch}$ & $0.01\div 0.07$& $0 \div 0.03$ & $1\div 2$ & $-$ & $0\div4$  \\
    \hline

    \end{tabular}
\caption{ Values of the model parameters obtained from the $SneIa + BAO + CMB$ data  for distance modulus versus theoretical results for the two-component fluid universe with  constant and varying $\Lambda$. }
  \label{tab:2Table}
\end{table}

\newpage
\section*{Discussion}
\begin{figure}[h!]
 \begin{center}$
 \begin{array}{cccc}
\includegraphics[width=80 mm]{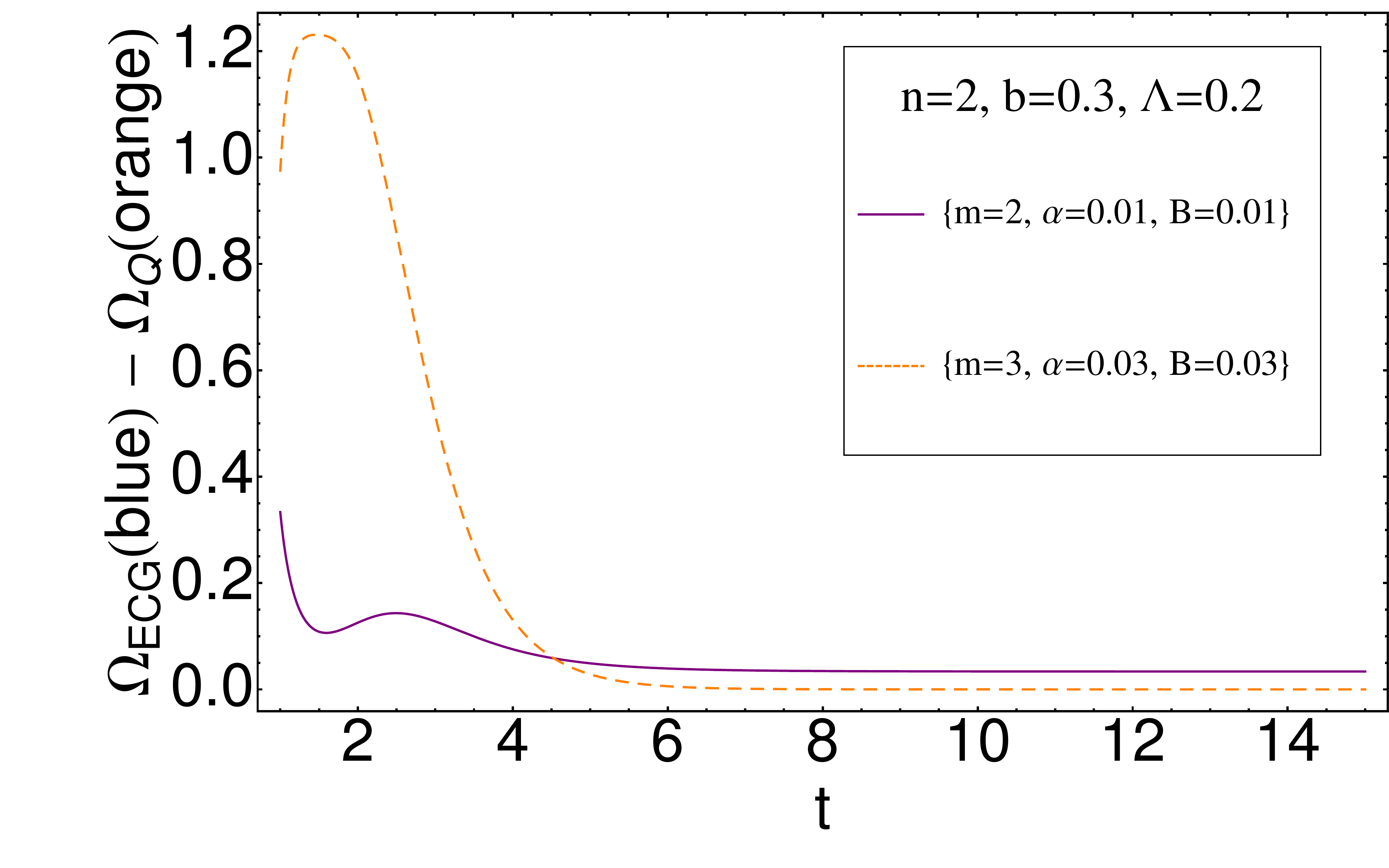} &
\includegraphics[width=80 mm]{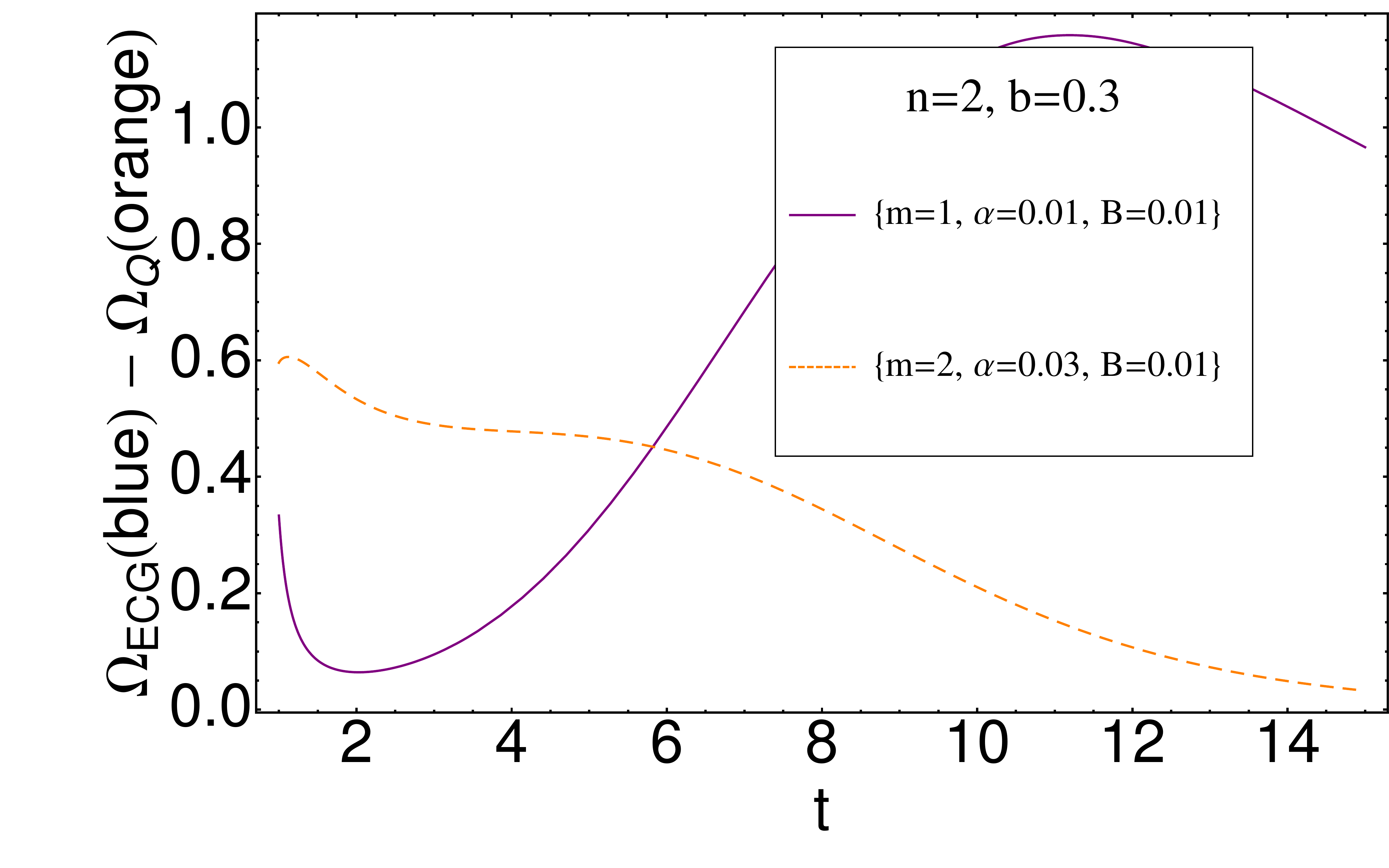}&
 \end{array}$
 \end{center}
\caption{Behavior of the critical densities $\Omega_{ECG}=\rho_{Ch}/3H^{2}$ and $\Omega_{Q}=\rho_{Q}/3H^{2}$ against $t$ for the constant $G$ and $\Lambda$ represents the left plot. The right plot represents the behavior of the critical densities $\Omega_{ECG}$ and $\Omega_{Q}$ assuming $\Lambda(t)=\rho_{Ch}+\rho_{Q}$ with the constant $G$.}
 \label{fig:modelOmega}
\end{figure}

In this Paper, the first cosmological model with the constant $\Lambda$ corresponds to the universe with  an extended Chaplygin gas as the content of it. The second model we considered to be an effective two-component fluid universe with the interacting extended Chaplygin gas and the scalar field quintessence DE. For the second model we perform analysis for two cases accounting  the constant and varying cosmological constants. Instead of field equations of GR theory, modified field equations mediated by Lyra geometry are used to describe the dynamics of the universe. The models are investigated in general manner for a class of self interacting potentials $V(\phi)$ of the exponential form. We found that all models are able to reproduce the accelerated expansion of the universe, where the content of the universe has a reasonable behavior. Behavior and the values of the Hubble parameter $H$ and the deceleration parameter $q$ are in the acceptable numerical range comparable with observations. In order to illuminate only the correct behavior of our models from the physical point of view we applied observational and causality constrains. Obtained constraints on the models parameters agree with the previously obtained results. We finish our work with the conclusion that we proposed and analyzed cosmological models with great accuracy and find out that our models are good models for the old universe. According to our models very old Universe corresponds to the de-Sitter Universe with $\omega=-1$, where extended Chaplygin gas will control the dynamics of the universe, while in the beginning quintessence DE would be dominating and controlling the dynamics of the universe (Fig~\ref{fig:modelOmega}).
We think that for the future analysis it will be interesting to include different forms of the interactions/couplings between extended Chaplygin gas and quintessence DE for this background, we also can investigate, for instance, the behavior of the models in the presence of the viscosity, which is known to be one of the ways to model the irreversible processes in an isotropic and homogeneous universe.


\newpage
\begin{thebibliography}{1}

\bibitem{Hu}
W. Hu and D. J. Eisenstein, Phys. Rev. D 59, 083509 (1999)

\bibitem{Hawking}
S. W. Hawking and G. F. R. Ellis, The large scale structure of space-time (Cambridge University Press, Cambridge, England, 1973)

\bibitem{Wald}
R. Wald, General Relativity (University of Chicago Press, Chicago, 1984)  

\bibitem{Erickson}

J.K. Erickson, R. R. Caldwell, P. J. Steinhardt, C. Armendariz-Picon, and V. F. Mukhanov, Phys. Rev. Lett. 88, 121301 (2002)

\bibitem{Ricardo}
Ricardo Garcia-Salcedo, Tame Gonzalez and Israel Quiros, Phys. Rev. D 89, 084047 (2014)

\bibitem{Riess}
A.G. Riess, et al., Astron. J. 116, 1009 (1998) 

\bibitem{Perlmutter}
S. Perlmutter, et al., Astrophys. J. 517, 565 (1999)

\bibitem{Bennett}
C.L. Bennett, et al., Astrophys. J. Suppl. 148, 1 (2003)

\bibitem{Spergel}
D.N. Spergel, et al., Astrophys. J. Suppl. 148, 175 (2003) 

\bibitem{Tegmark}
M. Tegmark, et al., Phys. Rev. D 69, 103501 (2004)

\bibitem{Abazajian}
K. Abazajian, et al., astro-ph/0410239 

\bibitem{Abazajian1}
K. Abazajian, et al., Astron. J. 128, 502 (2004)

\bibitem{Abazajian}
K. Abazajian, et al., Astron. J. 126, 2081 (2003) 

\bibitem{Hawkins}
E. Hawkins, et al., Mon. Not. Roy. Astron. Soc. 346, 78 (2003)

\bibitem{Verde}
L. Verde, et al., Mon. Not. Roy. Astron. Soc. 335, 432 (2002)

\bibitem{Kahya}
E.O. Kahya, B. Pourhassan, R. Myrzakulov, A. Pasqua, M. Khurshudyan, arXiv:1402.2592v2

\bibitem{Xu}
J. Lu, L. Xu, J. Li, B. Chang, Y. Gui, H. Liu, Phys. Lett. B 662, 87 (2008)

\bibitem{Toribio}
A. M. Velasquez-Toribio, M. L. Bedran, Brazilian Journal of Physics 41, 59 (2011)

\bibitem{Lu_Xu}
J. Lu, L. Xu, Y. Wu, M. Liu, Gen. Rel. Grav. 43, 819 (2011)

\bibitem{MKhurshudyan}
M. Khurshudyan, E. Chubaryan, B. Pourhassan, Int. J. Theor. Phys. 53 (2014) 

\bibitem{Guo}
Z. K. Guo, N. Ohta and Y. Z. Zhang, Mod. Phys. Lett. A 22, 883 (2007)

\bibitem{Dutta}
S. Dutta, E. N. Saridakis and R. J. Scherrer, Phys. Rev. D 79, 103005 (2009)

\bibitem{Saridakis}
E. N. Saridakis and S. V. Sushkov, Phys. Rev. D 81, 083510 (2010)

\bibitem{Peng}
Peng Wang, Puxun Wu, Hongwei Yu, Eur. Phys. J. C 72, 2245 (2012)

\bibitem{MKhurshudyan_2}
M. Khurshudyan, B. Pourhassan, R. Myrzakulov, S. Chattopadhyay, E.O Kahya, arXiv:1403.3768

\bibitem{Momeni}
Mubasher Jamil, Sajid Ali, D. Momeni, R. Myrzakulov, Eur. Phys. J. C 72 (2012)

\bibitem{Interaction_Saridakis}
Xi-ming Chen, Yungui Gong and Emmanuel N. Saridakis, arXiv:0812.1117v2

\bibitem{Ferreira}
P.G. Ferreira, M. Joyce, Phys. Rev. Lett. 79, 4740 (1997)

\bibitem{Copeland}
E.J. Copeland, M. Sami, S. Tsujikawa, Int. J. Mod. Phys. D 15, 1753 (2006)

\bibitem{Copeland1}
E. J. Copeland, A. R. Liddle and D. Wands, Phys. Rev. D 57, 4686 (1998)

\bibitem{Gong}
Y.G. Gong, A. Wang, Y.Z. Zhang, Phys. Lett. B 636, 286 (2006)

\bibitem{Lyra}
G. Lyra, Math. Z. 54, 52 (1951)

\bibitem{Sen}
D.K. Sen,  Z. Phys. 149, 311 (1957)

\bibitem{Dunn}
K.A. Dunn,  J. Math. Phys. 12, 578 (1971)

\bibitem{Halford}
W.D. Halford,  Aust. J. Phys. 23, 863 (1970)

\bibitem{Shchigolev}
V.K. Shchigolev, arXiv:1307.1866v1

\bibitem{Samanta}
G.C. Samanta, Int J Theor Phys 52, 3442 (2013) 

\bibitem{Martiros}
M. Khurshudyan, J. Sadeghi, R. Myrzakulov, A. Pasqua, H. Farahani, Adv.in HEP. 878092 (2014)  

\bibitem{Martiros1}
M. Khurshudyan, arXiv:1403.0109

\bibitem{Martiros2}
M. Khurshudyan, J. Sadeghi, A. Pasqua, S. Chattopadhyay, R. Myrzakulov, H. Farahani,  arXiv:1402.5678

\bibitem{Martiros3} 
M. Khurshudyan, A. Pasqua, J. Sadeghi, H. Farahani, arXiv:1402.0118


\end{thebibliography}
\end{document}